\documentclass{article}
\usepackage{amsmath}
\usepackage{amsfonts}
\usepackage{graphicx}
\usepackage{textcomp}

\DeclareMathOperator{\arsinh}{arsinh}
\DeclareMathOperator{\arcosh}{arcosh}
\DeclareMathOperator{\artgh}{artgh}
\DeclareMathOperator{\tgh}{tgh}
\DeclareMathOperator{\suc}{suc}
\DeclareMathOperator{\pre}{pre}
\DeclareMathOperator{\inv}{inv}
\DeclareMathOperator{\ladder}{ladder}
\DeclareMathOperator{\nxt}{nxt}
\DeclareMathOperator{\minus}{minus}

\DeclareMathOperator{\sqr}{sqr}

\begin{document}

\title{Criteria for the numerical constant recognition}

\author{A. Odrzywolek}

\maketitle

\begin{abstract}
The need for recognition/approximation of functions in terms of elementary functions/operations emerges in many areas
of experimental mathematics, numerical analysis, computer algebra systems, model building, machine learning,
approximation and data compression. One of the  most underestimated methods is the symbolic regression. In the article, reductionist 
 approach is applied, reducing full problem to constant functions, i.e, pure numbers (decimal, floating-point).  However, existing solutions are plagued by lack of solid criteria distinguishing between random formula, matching approximately or literally decimal expansion and probable ''exact'' (the best) expression match in the sense of Occam's razor. In particular, convincing STOP criteria for search were never developed. In the article, such a criteria, working in statistical sense, are provided. Recognition process can be viewed as (1) enumeration of all formulas in order of increasing Kolmogorov complexity K (2) random process with appropriate statistical distribution (3) compression of a decimal string. All three approaches are remarkably consistent, and provide
essentially the same limit for practical depth of search. Tested unique formulas count must not exceed 1/sigma, where sigma is relative numerical error of the target constant. Beyond that, further search is pointless, because, in the view of approach (1), number of equivalent expressions within error bounds grows exponentially; in view of (2), probability of random match approaches 1; in view of (3) compression ratio much smaller than 1.  
\end{abstract}

\begin{section}{Introduction}

Ability to recognize (approximate, identify, simplify, rewrite) function of several variables is a foundation of many scientific, model-building and machine learning techniques. Nowadays, deep neural networks are proven method \cite{backpropagation} to handle variety of practical problems, where some function $F(x,y,z,\ldots, a,b,c \ldots)$ of input variables $x,y,z, \ldots$ with parameters $a,b,c, \ldots$ must be created to approximate and generalize (usually big) data. Constants  $a,b,c$ are numerous weights found in the so-called training procedure \cite{backpropagation, ADAM}. However, this is not the only method to provide solution \cite{AIFeynmann} for unknown functions $F(x,y,z,\ldots, a,b,c \ldots)$. Another quite successful approach is the symbolic regression \cite{Eurega}, where sigmoidal neurons (aka layers) are generalized to any possible elementary function combinations 
and their compositions. To find particular combination of functions/operators of interest brute-force search or genetic methods could be applied.
To understand details of the symbolic regression procedure in the following research I maximally reduced original problem of finding function of several variables/parameters to $F(x,y,z, \ldots, a,b,c \ldots) = const$ case, i.e., decimal/numerical constants. However, it is important
to stress that discussed algorithms, can be used to identify functions as well \cite{SymbolicRegressionPackage}. But due to ,,curse of dimensionality'' original problem is much more complicated to analyze, and require tremendous computational resources as well. Therefore, to achieve some progress
and practical results, in the article I concentrate on the simplest possible case of $F \equiv const$. Example of $F=const$, is also a mandatory building block of more advanced symbolic regression techniques, where fitted parameters $a,b,c, \ldots$ must be identified with
some exact symbolic numbers, like combinations of $\pi$ and/or $e$, allowing for recursive search algorithm.

 However, identification of numerical constants itself is not an easy task \cite{BorweinBailey,RamanujanMachine,plouffe}. We are given decimal expansion only, and ask if there is some formula for it. There is strong demand from users of mathematical software to provide such a feature (identify/Maple \cite{Maple}, FindFormula/Mathematica \cite{Mathematica}, nsimplify/SymPy \cite{symPy}, RIES \cite{RIES} ). In typical situation we encounter some decimal number, resulting from numerical simulation or experiment, e.g: 1.82263,  and ask ourselves if it is equivalent to some symbolic expression, like e.g: $\sqrt{2}^{\sqrt{3}}$. Sometimes problem is trivial, for well-known numbers like 3.1415926 or 1.444667861. But in general, such a problem, without additional constraints, is ill-posed. Provided by the existing software \cite{Maple, symPy,RIES, Mathematica} ''answers'' nonsense. Often results are ridiculously complicated \cite{ISCwayback}. This is not a surprise, as Cardinality of real/complex transcendental numbers \cite{euler} is uncountably infinite (continuum $\mathfrak{c}$), while number of formulas and symbols is countable (aleph-zero $\aleph_0$). Therefore probability for randomly chosen real number to be equivalent to some formula is zero. Moreover, we must therefore restrict to numbers with finite decimal expansion. In practice, due to widespread of floating-point hardware, double precision \cite{double} being \textit{de facto} standard number of digits is quite small, around 16. There are (countable) infinity of
 formulas reproducing exactly those 16 digits, which numerically differ only at more distant decimal places: $1.8226300001, 1.8226300000009 ,1.822630000100017$ etc. But one still can ask \textbf{which one} of these formulas has \textit{lowest} Kolmogorov complexity $K$ \cite{complexity}. To quantify value of $K$, we must specify ''programming language'' for generating formulas.

 Above considerations lead to the following, well-posed, formulation of the numerical constant recognition problem. Imagine a person with hand-held
 scientific calculator, who secretly push a sequence of buttons, and print out the decimal/numerical result.
 We ask if this process can be reversed?  In other words, given numerical result, are we able to recover sequence of calculator buttons? 
 Answer to this question is the main goal of the article.

 Since every meaningful combination of buttons is equivalent to some explicit, closed-form \cite{closed}, mathematical formula, above thought experiment becomes practical formulation of the constant recognition problem. In a real world implementation, human and calculator are replaced by the computer program\footnote{Alternatively, database of the pre-computed results \cite{ISCwayback}.}, increasing speed billion-fold. Mathematically,
results which can be explicitly computed using hand-held scientific calculator are members of so-called
$\exp-\log$ (EL) numbers \cite{chow}. Above formulation has many advantages over set-theoretic or decimal string matching approach. It is well-posed, tractable, and can be either answered:\\
a) positively, \\
b) in terms of probability, or\\ 
c) falsified.\\

Precise answer depend on both maximum length of button sequence (code length, Kolmogorov complexity, denoted by $K$), as well as precision of the numerical result. One could anticipate, that for high-precision numbers (arbitrary precision in particular) and fixed, short code (calc button sequence length), identification will be unambiguous. For intermediate case (e.g. double precision, unspecified code length) we expect to provide some probability measure for identification and/or ranked list of candidates. For low-accuracy numbers (e.g. of experimental origin) failure seems inevitable. Article is devoted to quantify above considerations. Certain statistical criteria are presented for practical number identification. Hopefully, this will convince researchers, that constant recognition problem can be solved practically. Instead of software used mainly for recreational mathematics and random guessing, it could become reliable tool for numerical analysis and computer algebra systems.

Article is organized as follows. In Sect.~\ref{ELnumbers} we discuss simplified computer language (in replacement for physical scientific calculator) used to define Kolmogorov complexity of the formulas. In Sect.~\ref{enum} we discuss some practical issues regarding numerical implementation of the sequential formula generators.  Then, we combine experience/knowledge of Sect. \ref{ELnumbers} and \ref{enum} to find properties and typical behavior of the constructed EL numbers. Sections \ref{Criterion1}-\ref{Criterion3} propose three criteria for number recognition: (1) instant error drop-off to machine epsilon instead statistically expected $e$-folding; exponential growth of number of formulas indicates failure of search (2) maximum likelihood formula in the view of statistical process (3) compression ratio of the decimal constant in terms of RPN calculator code string.

\end{section}

\begin{section}{Elementary complex exp-log numbers \label{ELnumbers}}

The first task is to precisely define set of formulas/numbers (calculator used) we want to identify
using decimal expansion. I restrict myself to explicit formulas \cite{chow}, i.e., those computable
using hand-held scientific calculator. We assume root-finding procedure is absent, i.e., formula must be identified
in explicit form, not if form of an equation \cite{RIES} or in implicit form. Therefore, as notable exception, 
most of algebraic numbers are not in above class, because polynomials of the order
5 and larger are not explicitly solvable\footnote{But they might be identified in the above sense if we expand class of the functions including e.g.  elliptic functions.} in general. To be specific, I consider
any complex number created from
\begin{enumerate}
\item integers: $0,1,-1,2,-2, \ldots$
\item rationals: $\frac{1}{2},\frac{1}{3}, \frac{3}{2}, \ldots$
\item mathematical constants: $0, 1, 2, e,\pi, i, \phi, \ldots$
\item addition/subtraction, multiplication/division, exponentiation/logarithm (arbitrary base)
\item elementary functions of one variable:
\begin{itemize}
\item trigonometric: $\sin, \cos, \tan$ 
\item inverse~trigonometric (cyclometric): $\arcsin, \arccos, \arctan$
\item hyperbolic: $\sinh, \cosh, \arsinh, \arcosh, \tgh, \artgh$
\item $\sqrt{\quad}  \;, \; \lg_2 \; , \;  \ln \;, \; \exp$
\item usually unnamed functions: $1/x, x+1, -x, x^2, 2x, x^x \ldots$
\end{itemize}  
\item function composition.
\end{enumerate}

The choice of complex instead of real field is justified by simplicity. Intermediate results are always valid, except division by zero. In practical
numerical implementation choice of complexes might be sub-optimal, due to reduced numerical performance and still missing
implementation of some elementary functions (e.g: \texttt{clog2}) in standard libraries \cite{glibc}. 
Last item in the above list, function composition, is important for completeness. E.g, $\ln{(\ln{(\ln{x})})}$ is well-defined
elementary function, although rarely used in practice. 
No special functions were used, but it is easy to include them \cite{SymbolicRegressionPackage} in Computer Algebra System (CAS) environment. 
Inside CAS one can use symbol $\infty$, and high-level mathematical constructs like sums \cite{RamanujanMachine}, derivatives and integrals . Here I concentrate mainly on elementary functions, to utilize low-level programming and hardware-accelerated calculations.

For beginners in the field, troublesome are symbols (buttons) related to integers and rationals. Sets of integers and rationals are infinite. Without proper handling enumeration quickly will get stuck on rational
approximations with very large integer numerator/denominator \cite{Mathematica}. Or worse, skip some of them. Therefore, we must restrict to some ''small'' subset integers and rationals. Handheld calculators have ten digit buttons $0123456789$ and use positional system to enter integers. 
Standard computer languages (C, C++, Fortran)  use ''small'' integers still too large from our point of view, e.g. 8-bit
signed and unsigned ones. Complexity of numbers 0, $2^{7}=128$, 188, \ldots is identical
using such an approach. The other extreme follows reductionist definition of \cite{Tarski}, staring from e.g: $-1$, and constructing all other integers as follows:
$$
1 = (-1) \times (-1), \; 0 = 1 + (-1), \; 2=1+1, \ldots
$$
In calculators, as we mentioned, ten digits are present: 0,1,2,3,4,5,6,7,8,9 and integers are entered
in sequence using standard positional numeral system. RIES \cite{RIES} follow similar way, except for 0 and 1, which are missing. Use of positional base-10 numerals is simple
reflection of human anatomy/history, but difficult to justify and include in our virtual calculator. Therefore we restrict to a few small magnitude integers:  $-1,0,1,2$. All other
integers must be constructed from above four. Noteworthy, above four integers are also among
essential mathematical constants included in famous \cite{e} Euler formula:
\begin{equation}
\label{euler}
e^{i \pi} + 1 = 0, \qquad e^{2 \pi i} - 1 = 0.
\end{equation}
Inclusion of 2 is required, because it is the smallest possible integer base for logarithm $\log_2$. Last but not least, binary system
is now used in virtually all modern computer hardware.

The same reasoning applies to rationals. One might enumerate them using Cantor diagonal method, Stern-Brocot tree, or generate
unambiguously by repeated composition\footnote{See also Appendix~B in Supplementary Material.} of a function:
$$
\nxt(x) = \left(1 + 2 \lfloor x \rfloor -x  \right)^{-1},
$$
starting from zero: 
$$
0,1,\frac{1}{2},2,\frac{1}{3},\frac{3}{2},\frac{2}{3},3,\frac{1}{4},\frac{4}{3},\frac
   {3}{5},\frac{5}{2},\frac{2}{5}, \ldots
$$
Usually the best is to leave language without explicit rationals, and let them appear by division
of integers. Alternatively, addition of reciprocal $\inv(x) \equiv 1/x$ and successor $\suc(x) \equiv x+1$ functions allow for construction of continued fractions, see Appendix~B in Supplementary Material.

Set of (named) transcendental mathematical constant certainly must include $e$ and $\pi$. Imaginary unit $i = \sqrt{-1}$, 
due to \eqref{euler}, is required as long as we consider complex numbers (I do). Besides this, some mathematicians consider other constants \cite{const} as ''important'': small square roots ($\sqrt{2}, \sqrt{3}, \sqrt{5}$), golden ratio $\phi = (1+\sqrt{5})/2$ \cite{RIES}, $\ln{2}$, $e^{1/e}$ and so on. Noteworthy, all of them are itself in $\exp-\log$ class, i.e., we can generate them from integers and elementary functions. However, inclusion of them alter language definition (size) and therefore Kolmogorov complexity. Unclear situation is with inclusion of mathematical constants of unknown status/memberships (unknown to be rational or transcendental), like Euler gamma $\gamma = 0.577216\ldots $ or Glaisher constant $A=1.28243\ldots$. They may, or may not, extend $\exp-\log$ class. Exploration of them is itself a major application for constant recognition software \cite{RamanujanMachine}.

Commutative operations like addition and multiplication are in general $n$-ary. To simulate calculator
behavior we must treat them as repeated binary operations
$$
x+y+z = (x+y) + z.
$$

Elementary functions listed above, at the beginning of this Section,  are not independent. For example
$$
\sinh{x} = \frac{e^x - e^{-x}}{2}, \tan{x} = \frac{\sin{x}}{\cos{x}}.
$$
Scientific calculators sometimes distinguish less important, secondary functions
which require pressing two buttons. But mathematically
they are completely equivalent. You can compute $\sinh, \cosh$ using $\exp$, or \textit{vice versa}.
This is especially true in complex domain, where all elementary functions are reducible to $\exp$
and $\ln$.

Reduction is possible for binary operations as well, e.g:
$$
x \times y = e^{\ln{x} + \ln{y}}, \quad   x \cdot y = \log_x{\left[ (x^y)^x \right] }.
$$
Replacing multiplication by logarithms and addition is an achievement of medieval 
mathematics \cite{briggs} used without changes for 300 years. It was replaced with slide rule used in XX century.  
They both went extinct with modern computers. However, we point out, that using logarithms and addition you
can not only do bottom-up translation to multiplication/exponentiation. You might go opposite way as well, from high-rank
(grade) hyper-operation down do additions. Expressing addition/multiplication by exponentiation/logarithms only (up-bottom) is therefore possible, but tricky, see Appendix~C in Supplementary Material.

In the middle of the above considerations, intriguing question arises: how many constants, functions and binary operations are required for our virtual calculator to be still fully operational? How far reduction process can go, without impairing our computational abilities? This is also known
as the ,,broken calculator problem'' \cite{RIES}. Another related question
is, if such a reduced calculator is optimal for the task of constant recognition. Obviously,
maximally reduced button set makes theoretical and statistical analysis convenient, therefore
it is very useful as mathematical model. In practice, as I will show later, Kolmogorov complexity of formulas in maximally reduced language, 
which are considered as simple by humans, might become surprisingly large. On the contrary, formulas \textit{simple} in reduced operation set, like
nested power-towers, look inhuman, and are out of scope traditional mathematical aesthetics, despite small Kolmogorov complexity.

The simplest possible language I found so far has a length three ($n=3$). It is still able to perform all operations
of the scientific calculator. It includes:
\begin{subequations}
\label{base-3}
\begin{align}
\label{e}
\text{either} & \qquad e \quad  \text{or}  \quad \pi. 
\end{align}
\begin{align}
\text{binary exponentiation} & \qquad x^y,
\end{align}
\begin{align}
\text{arbitrary base (two-argument) logarithm} &\qquad  \log_x{y}.
\end{align}
\end{subequations}

Another very simple language of length $n=4$ is:
\begin{subequations}
\label{base-4}
\begin{align}
\text{any constant}& \qquad x.
\end{align}
\begin{align}
\label{e1}
\text{natural exponential function}& \qquad \exp{x}  \equiv e^x,
\end{align}
\begin{align}
\label{e2}
\text{natural logarithm}& \qquad \ln{x}  \equiv \log_e{x},
\end{align}
\begin{align}
\text{subtraction}& \qquad x-y.
\end{align}
\end{subequations}

\begin{figure}
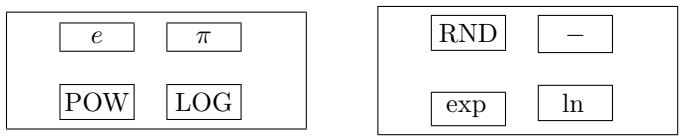

\framebox[4cm]{
\parbox{3cm}{
\begin{tabular}{cc}
\framebox[1cm]{$e$} & \framebox[1cm]{$\pi$} \\
&\\
\framebox[1cm]{POW} & \framebox[1cm]{LOG} 
\end{tabular}
}
}
$\qquad$
\framebox[4cm]{
\parbox{3cm}{
\begin{tabular}{cc}
\framebox[1cm]{RND} & \framebox[1cm]{$-\vphantom{\text{RND}}$} \\
&\\
\framebox[1cm]{$\exp{}$} & \framebox[1cm]{$\ln{}$} 
\end{tabular}
}
}
\caption{\label{4-button-calc} Graphical representation of the two simplest, fully operational hand-held RPN calculators. 
Up-bottom CALC1 \eqref{base-3} is on the left. Bottom-up CALC2 \eqref{base-4} on the right. Descriptions reflect rank of available
binary (hyper-)operation: exponentiation (rank 3) on the left, and subtraction (rank 1) on the right. Calc on the left in fact
require only three buttons, either $e$ or $\pi$ is required. Since calc on the right require
any complex number button, it was replaced with random value. In practice it will
be some small integer, like $-1,0,1,2$, but can be anything, e.g, $i$ or Euler $\gamma$. See also Fig.~\ref{mma}.}
\end{figure}

Detailed proofs by exhaustion are presented in Appendices C,D in Supplementary Material. 
Above two examples clearly justify name $\exp-\log$ for class of explicit elementary numbers discussed
in this section. I was unable to find any shorter languages. At least one (noncommutative?) binary operation seem required to start abstract syntax (AST) tree growth (see however Appendix~B), and at least one constant to terminate leafs. One of the essential constants
$e$ or $\pi$ probably also is required, either explicitly, like in \eqref{e}, or hidden in function/operation definition
(\ref{e1},\ref{e2}). But I cannot prove that further reduction to 2 buttons is impossible. I'm unable to show, that above language 
of length three \eqref{base-3} is unique. Therefore, one cannot estimate
Kolmogorov complexity of EL number unambigously. Existence of both smallest and unique language,
generating all $\exp-\log$ numbers, would provide natural enumeration of mathematical formulas, similar
to Peano arithmetic for natural numbers. Implications of such a discovery would be tremendous. For example, any
mathematical formula could be linked with unique ''serial number'' and easily searched on the internet. Anyway, \eqref{base-3} or \eqref{base-4} are
the sets of irreducible operations, used for now. Calculator/language can be extended, but under no circumstances any of the buttons defined by \eqref{base-3} or \eqref{base-4} can be removed. Failure to abide by above requirement might result in catastrophic failure of the number recognition software even in the simplest test cases. This is plague
of existing implementations, letting unaware end-users to choose base building blocks on their own. This leaves impression of random failures without any obvious reason, discouraging users and possible applications. Users can use their own sets of constants, functions and binary operations, but they must be merged with irreducible set \eqref{base-3} or \eqref{base-4}. This is a major result in the field of symbolic regression, and can be 
very easily overlooked for general functions of several variables. Irreducible, minimal calculators are depicted visually in Fig.~\ref{4-button-calc}.

It is illustrative to compare \eqref{base-3} and \eqref{base-4} with \textit{Mathematica} core language \cite{Mathematica}, composed of:
\begin{subequations}
\label{mma}
\begin{align}
\text{addition (\pmb{Plus})} & \qquad x+y,
\end{align}
\begin{align}
\text{multiplication (\pmb{Times})} & \qquad x \times y,
\end{align}
\begin{align}
\text{exponentiation (\pmb{Power})} & \qquad x^y,
\end{align}
\begin{align}
\text{natural logarithm (\pmb{Log})} &\qquad  \ln{x},
\end{align}
\begin{align}
\text{mathematical constants (\pmb{E}, \pmb{Pi}, \pmb{I})} \qquad e, \pi, i, \ldots
\end{align}
\end{subequations}
augmented with arbitrarily large integers, including
their complex combinations (\pmb{Complex}) and rationals (\pmb{Rational}). Visual comparison of \eqref{base-3}, \eqref{base-4} and \eqref{mma} is presented in Fig.~\ref{BaseSet}. 

\begin{figure}
\includegraphics[width=0.69220062755534635386542199718279\textwidth]{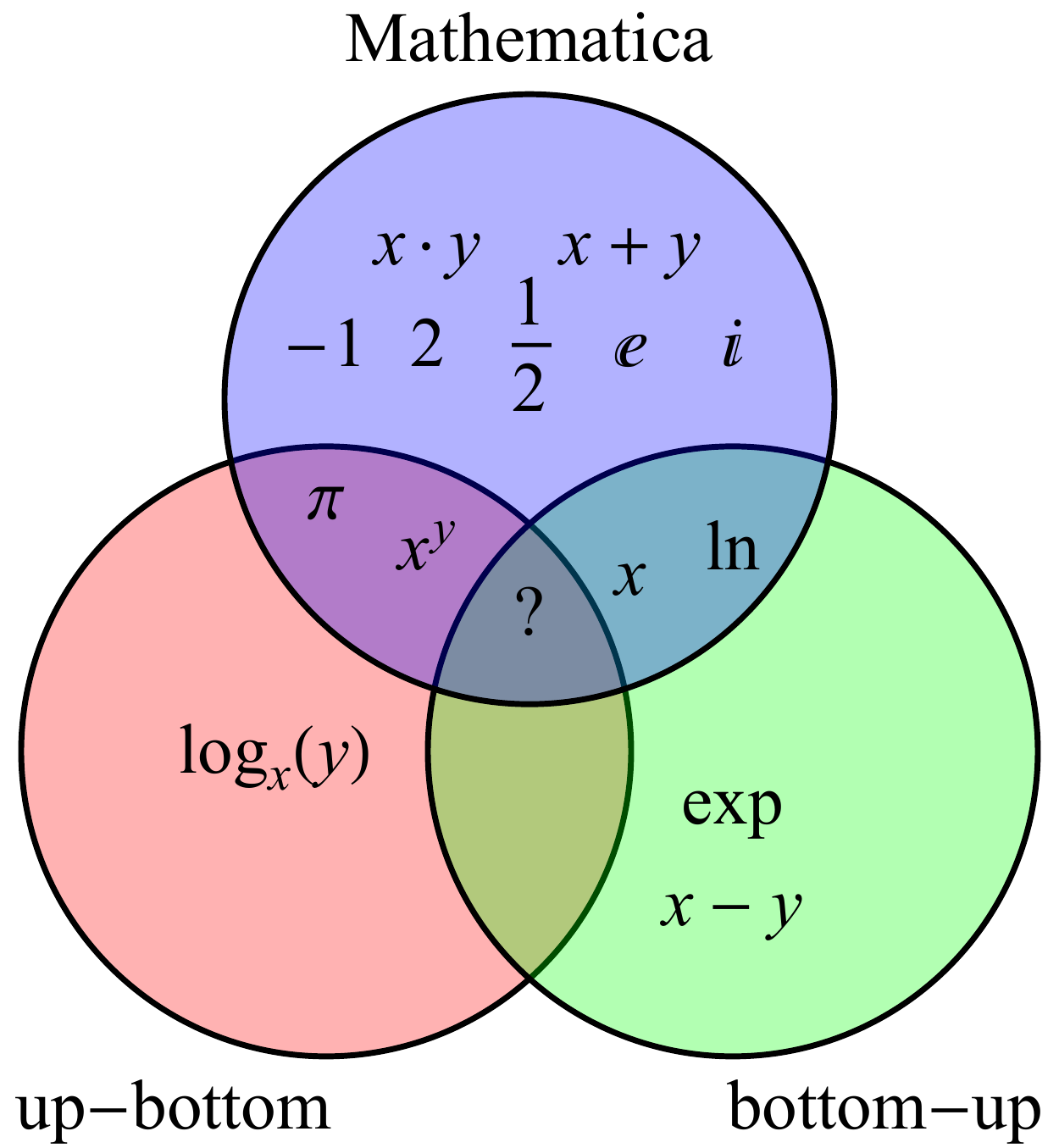}
\caption{\label{BaseSet} Illustration of base constants, functions and binary operations used to define
$\exp-\log$ number class. Irreducible set \eqref{base-3} is labeled ''up-botton'', because it is based
on highest rank complex (hyper-)operation currently implemented (exponentiation), while \eqref{base-4} (bottom-up) starts with low-rank arithmetic (subtraction).  \textit{Mathematica} core language is included for comparison. All three sets allow generation of any explicit elementary transcendental constant in the EL class. Question mark symbolize unknown ''holy-grail'' complex function, which could, if exists, by repeated composition, generate all $\exp-\log$ numbers in  unique order.
}
\end{figure}

Accounting for the above considerations, for further analysis we have selected four base sets (buttons) to attack the inverse calculator problem:
\begin{subequations}
\label{CALC}
\begin{enumerate}
\item \texttt{E}, \texttt{POW}, \texttt{LOG}     [$n=3$ buttons, eq.~\eqref{base-3} ] 
\begin{equation}
\label{CALC1}
\end{equation}
\item \texttt{X}, \texttt{EXP}, \texttt{LN}, \texttt{-}    [$n=4$ buttons, eq.~\eqref{base-4}]
\begin{equation}
\label{CALC2}
\end{equation}
\item \texttt{Pi}, \texttt{E}, \texttt{I}, \texttt{-1},
\texttt{2}, \texttt{1/2},  \texttt{Log}, \texttt{Plus}, \texttt{Times}, \texttt{Power} 
[$n=10$ buttons, eq.~\eqref{mma} ] 
\begin{equation}
\label{CALC3}
\end{equation}
\item full scale scientific calc [$n=36$ buttons, see below] 
\begin{equation}
\label{CALC4}
\end{equation}
\end{enumerate}
\end{subequations}

Calculator 1 (up-bottom), based on \eqref{base-3} is the simplest found. It use one constant ($e\simeq 2.71828\ldots$, PUSH on stack operation), no functions, and only two non-commutative binary operations: arbitrary base logarithm $\log_x{y}$ and exponentiation $y^x$. Noteworthy, AST (Abstract Syntax Tree) is a binary tree in case of \eqref{CALC1}.

Calculator 2 (bottom-up), based on \eqref{base-4} is the second shortest, and have remarkable
property: it can use any numerical constant/symbol $x$ to generate all other numbers. For example,
$0 = x - x$, $1 = \exp{(0)}$, $e=\exp{(1)}$ and so on.  We can use any $x$:  $x=0$, $x=-1$, $x=e$, $x=i$, $x=\phi$, \ldots Even a target
constant $z$ being searched can be used!  This property can be exploited in many ways: (1) use our calculator to operate on vectors, (2) use $x$ with large Kolmogorov complexity to ''shift'' formula generator forward (3) use $x=z$ and search in implicit form. The first one can be used
to take advantage of modern CPU AVX extensions, the second to use $x$ as some form of random seed
for enumeration procedure. This can be done extending any of calculators \eqref{CALC} with $x$ as well, but only for \eqref{base-4} it is visible \textit{explicite} in definition. In presentation of results we used $x=2$. 

Calculator 3 is designed to mimic \textit{Mathematica} behavior. Constants beyond 
$\pi,e$ and $i$ were chosen as follows. To enable rapid integer generation via addition,
multiplication and exponentiation we use -1 and 2. Incidentally, these numbers are also are initial values for Lucas numbers. Rational numbers are easily generated via continued fractions, thanks to inclusion of $-1$ constant, allowing for construction of reciprocal $1/x \equiv x^{-1}$. Rational constant $1/2$ form square root $\sqrt{x} = x^{1/2}$, and together with imaginary magnitude $i$ generate trigonometric functions via Euler formula $e^{ix} = \cos{x} + i \sin{x}$. Number of instructions equal to ten is selected intentionally. It will allow for instant estimate of the compression factor in Sect.~\ref{Criterion3}. Both target numerical constant and RPN calculator code can be expressed as string of base-10 digits \texttt{0123456789}, see next Sect.~\ref{enum}.

Fourth calculator is the largest one. Maximum number of 36 buttons is limited by a current implementation of the fast \texttt{itoa} function used to convert string variables into base-36 numbers, i.e. alphanumeric lowercase digits and letters. It is the most close to what people usually expect from scientific hand-held device. Full list of buttons:
\begin{itemize}
\item constants: $1,2,3,4,5,6,7,8,9,e,\pi,i,\phi$
\item functions:$\ln, \exp, \inv, \minus, \mathrm{sqrt}, \sqr,$ 
$\sin, \arcsin, \cos, \arccos, \tan, \arctan,$ 
   $\sinh, \arsinh, \cosh, \arcosh, \tgh, \artgh,$
\item binary operations: $+, -, \times, / , y^x$.
\end{itemize}

''Full'' calculator defined above use 13 constants, 18 functions of one variable, and 5 binary operations.

\end{section}

\begin{section}{Efficient formula enumeration  \label{enum}}

Once (\textit{virtual}) calculator buttons, equivalent to some truncated computer language specification, are fixed, we face the next task: enumeration of all possible formulas. We exclude recursive implementation. While short and elegant,
it quickly consumes available memory, and is hard to parallelize \cite{RIES}. For sequential generation, we notice that formulas are equivalent to set of all abstract syntax trees (AST) or valid reverse polish \cite{RPNorig} notation (RPN, \cite{RPN}) calculator codes. The former description has an advantage in case where efficient tree enumeration algorithm exists \cite{Knuth}, e.g: binary trees \cite{tree}. Unfortunately, this is applicable only to the simplest calculator \eqref{CALC1}. Other mix binary and unary trees. Therefore, to handle variety of possible calculators, including future extension
to genuine $n$-ary special functions (e.g: hypergeometric, Meijer G, Painleve transcendents, Heun)
our method of choice is enumeration of RPN codes. This has two major advantages. First, enumeration is trivially provided by the standard \texttt{itoa} function with base-$n$ numbers (including leading zeros), where $n$ is the number of buttons. Second, while majority of enumerated codes is invalid, checking RPN syntax is very fast, almost negligible compared to \texttt{itoa} itself, let alone to computation of complex exponential and logarithmic functions.  Procedure has two loops: outer for code length $K$,
inner enumerating $n^K$ codes of length $K$. Inner loop is trivial to parallelize, e.g. using OpenMP \cite{OMP} directives without effort, scaling linearly with number of physical cores. Moreover, hyper-threading is utilized as well, although scaling is only at quarter of core count. Therefore, prospects for high-utilization of modern high-end multi-core CPU's \cite{epyc} are looking good.  
Digits are associated with RPN calculator buttons. For simplest case \eqref{base-3} they are ternary base digits \texttt{012}, which were assigned to three RPN
calculator buttons as: 0 $\to$ E ($e$), 1 $\to$ LOG ($\log_x{y}$), 2 $\to$ POW ($x^y$). Detailed example of the algorithm
for \eqref{base-3} is presented in Appendix~A of Supplementary Material.

Combinatorial growth of the enumeration is characterized, in addition to Kolmogorov complexity 
(RPN code length) $K$, by a total number of possible codes tested so far $k_1$. It grows with $K$ as:
$$
k_1 = \sum_{K'=1}^{K} n^{K'} = \frac{n^K-1}{n-1} n.
$$
Total number of syntactically correct codes
$k_2$ and total number of \textit{unique} numbers $k_3$ are additonal useful characterizations of the search depth. Obviously, $k_3 \leq k_2 \leq k_1$.
For perfectly efficient enumeration $k_1=k_2=k_3$. Unfortunately, this is currently possible only for rational numbers, see Appendix~B. Case $k_1=k_2$ is equivalent to knowledge of unique tree enumeration algorithm. To achieve $k_2=k_3$ ($p=1$ in Fig.~\ref{k2k3}, dashed diagonal line) algorithm must magically somehow know in advance all possible mathematical simplifications. Not only those trivial, like $e^{ \ln{x} } = \ln{(e^x)}=x$ but anything mathematically imaginable, e.g, $2^{1-\ln{\ln{\pi}}} = 2 (\ln{\pi})^{-\ln{2}}$. I doubt this is possible at all. However, some ideas, based on solved rational numbers enumeration, are presented in Appendix~B. In practice, we achieved  $k_3/k_2 = 0.59, 0.06, 0.08, 0.49$ for calculators 1-4 \eqref{CALC}, respectively. Large $k_3/k_2$ ratio for \eqref{CALC1} is a result of very simple tree
structure, in which only every nine-th odd-$K$ RPN codes are valid. Without taking this into account, $k_3/k_2=0.59/2/9 = 0.03$,
i.e., the worst of four. Full 36-button calculator perform surprisingly well. It appears that some economy and/or experience gained 
by centuries is behind it. 

\begin{figure}
\includegraphics[width=0.69220062755534635386542199718279\textwidth]{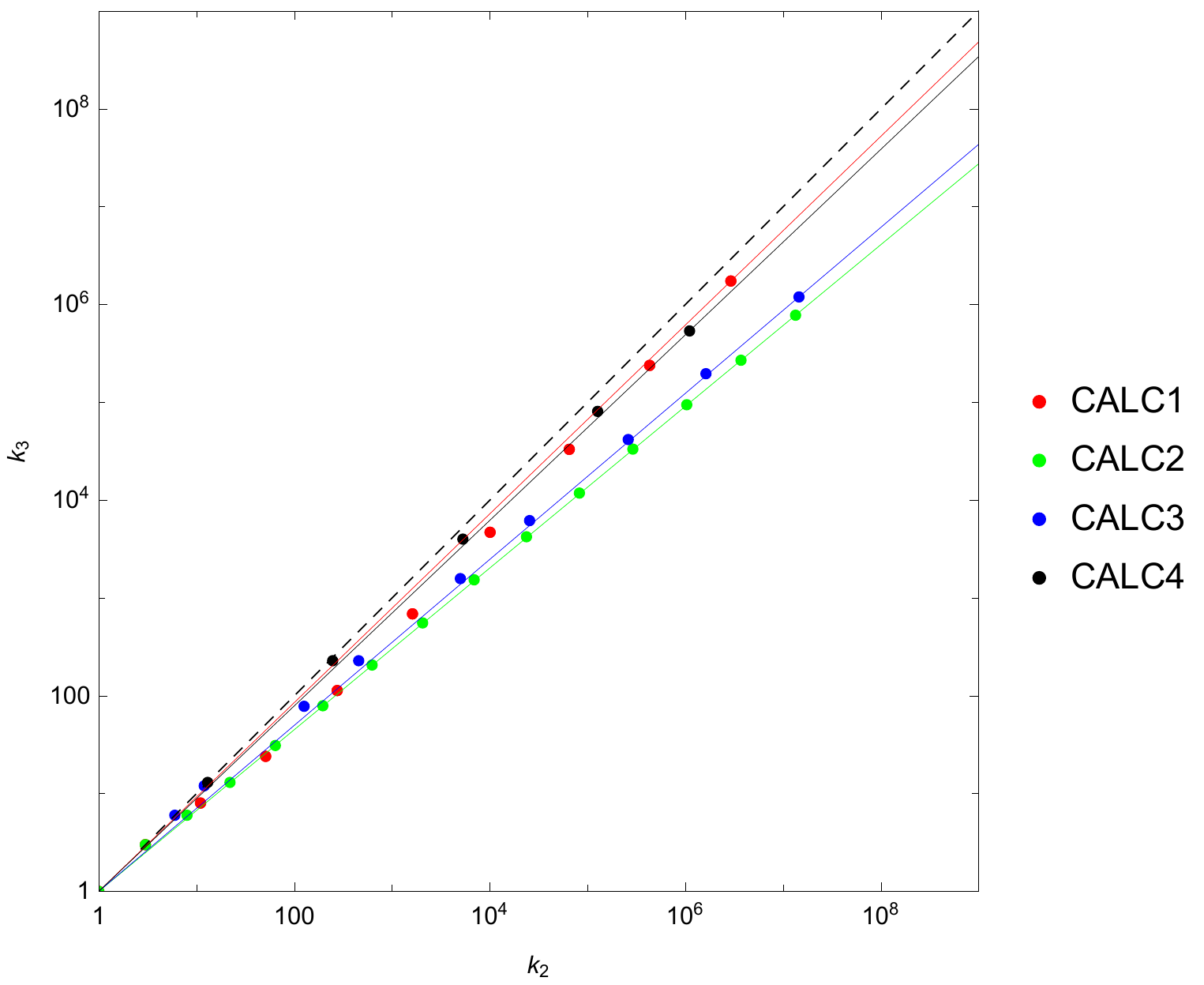}
\caption{\label{k2k3} Number of unique formulas $k_3$ \textit{versus} number of valid codes $k_2$. Solid lines show $k_3 = {k_2}^p$ fits, 
 with $p=0.96491, 0.826455, 0.848559, 0.948259$ for calculators 1-4, respectively. Dashed diagonal line shows perfect case $k_3=k_2$.  
}
\end{figure}

\end{section}

\begin{section}{Quasi-convergence rate of enumerated formulas \label{Criterion1} }

Enumerated formulas provide sequence of elementary transcendental $\exp-\log$ (EL) numbers. For \eqref{base-3}
they are, in order of increasing Kolmogorov complexity $K$:
\begin{align}
\label{ELseq}
K=1&\qquad e,                    \\
K=2&\qquad \text{none},   \nonumber        \\ 
K=3&\qquad 1, e^e,        \nonumber         \\
K=4&\qquad \text{none},    \nonumber         \\
K=5&\qquad 0, 1/e, e^{e^e}, e^{e^2}, \nonumber \\
\ldots  \nonumber
\end{align}
where repeated numbers were omitted. Position within the same level of $K$ is unspecified, but follows
from presumed enumeration loop and base-$n$ digits association with buttons. From sequence \eqref{ELseq} we can extract subsequence of progressively better approximations
for target number $z$, e.g., using example from the introduction, $z=1.8226346549662422143937682155941\ldots$. Analysis of sequences obtained this way is main goal of current section. We are interested in convergence properties and criteria for termination of sequence.

Let us assume for the moment, that we are able to provide on demand as many decimal digits for target number $z$ as required.
There are two mutually exclusive possibilities: (i) number $z$ is in $\exp-\log$ class, defined e.g. by \eqref{base-3}
or \eqref{base-4}, or (ii) $z$ is true non-elementary transcendental constant outside EL class. In the former case, we expect
that error eventually, at some finite $K$, will drop to an infinitesimally small value, limited only by currently used numerical precision. Search algorithm
could then terminate and switch to some high-precision or symbolic algebra verification method. Unfortunately, equivalence problem is undecidable in general \cite{richardson}. One cannot exclude, that both numbers deviate at some far more distant decimal place. In the case of (ii) i.e. truly transcendental number, however, sequence will converge indefinitely, by generation of approximations, with progressively more complicated formulas. Realistic convergence examples, computed using extended precision (\texttt{long double} in C) to prevent round-off errors, are presented in Figure~\ref{SmokingGun}. The example target $z=\sqrt{2}^{\sqrt{3}}$ from introduction was selected. Number is obviously in $\exp-\log$ class. But is not \textit{explicite} listed in any of \eqref{CALC}, so it must be entered using appropriate sequence of buttons. For RPN calculator \eqref{CALC4} it is obvioulsy 2, SQRT, 3, SQRT, POW with $K=5$. In Fig.~\ref{SmokingGun} is is visible at $K=5.28$, where decimal part show inner loop progress. 

\begin{figure}
\includegraphics[width=0.69220062755534635386542199718279\textwidth]{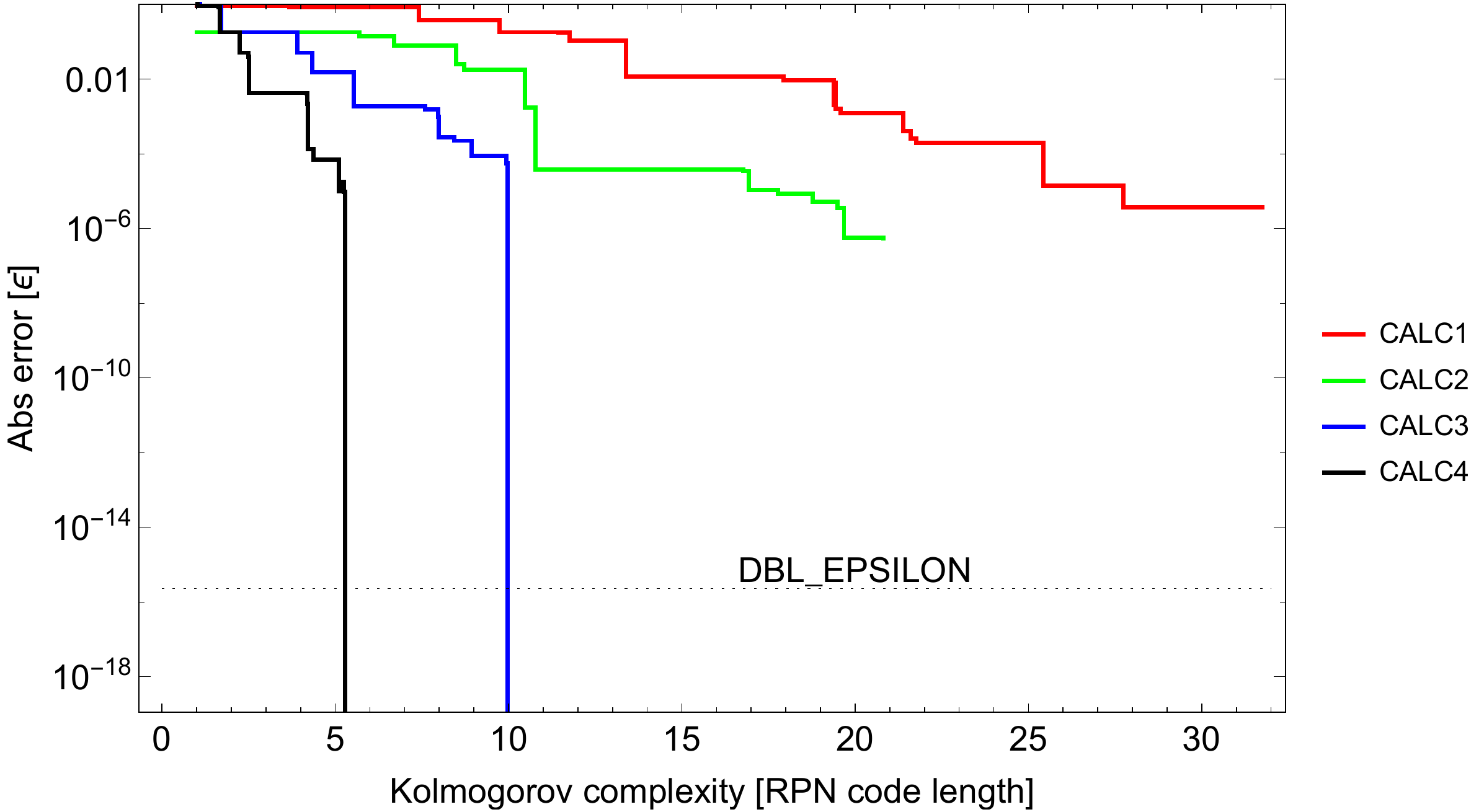}
\caption{\label{SmokingGun} Typical behavior of absolute error $\epsilon$ for subsequence of progressively better approximations
in terms of $\exp-\log$ formulas generated with ''calculators'' \eqref{CALC}, as a function of Kolmogorov complexity $K$. For \eqref{CALC2} we have used $x=2$.  For \eqref{CALC3} (blue) and \eqref{CALC4} (black) sudden error drop-off to machine epsilon (zero) is observed for small $K$, marking possible ''exact'' formula match. Calculations were done using extended precision (\texttt{long double} in C) while horizontal line mark machine epsilon for IEEE 754-2008 (binary64).
}
\end{figure}

Two types of behavior can be observed in Fig.~\ref{SmokingGun}. Vertical absolute error $\epsilon$ scale
is adjusted to range $\{ 2^{-63}, 1 \}$, i.e., \texttt{long double} machine epsilon for this
and related Figs. \ref{Convergence2}, \ref{Convergence3}. In typical situation absolute error $\epsilon$ for
best approximation decreases exponentially with code length. However, if ''true'' $\exp-\log$
formula is encountered, error instantly drops off to limiting value. In example from Fig.~\ref{SmokingGun}
it is small multiple of machine epsilon for \texttt{long double} precision ($2^{-63}$) or binary zero.

Criterion directly based on Kolmogorov complexity is inconvenient, if one deals with languages
of various size, like our calculators 1-4, eq.~(\ref{CALC}). Convergence rate also depends
on language size $n$, cf. Fig.~\ref{SmokingGun}. One could use Kolmogorov complexity
corrected for language size, or compile formulas generated in extended calculators \eqref{CALC3} or 
\eqref{CALC4} down to one of primitive forms given by \eqref{base-3} or \eqref{base-4}. The former approach usually underestimate, and the latter heavily overestimates true complexity. Therefore we propose another criterion, independent of the language used to generate formulas. Instead of plotting error as a function complexity,
we plot $N$-th best approximation (Fig.~\ref{Convergence2}). Now all curves nearly overlap, and
observed lower error limit is $\epsilon > e^{-N}$.

Above considerations provide the first criterion for constant identification:
\begin{center}
Criterion 1:\\ 

Identification candidate:
\textit{if absolute error $\epsilon$ in sequence \eqref{ELseq} of the progressively better approximations in terms
of the $\exp-\log$ formulas deviates ''significantly'' from estimated upper limit $e^{-N}$ (drops to 
numerical ''zero''/machine epsilon in particular) we can stop search and return formula code, as possible
identification candidate.}

Failure of search:
\textit{if absolute error in subsequence of progressively better approximations in terms
of $\exp-\log$ formulas follow $e^{-N}$ (dot-dashed line in Fig.~\ref{Convergence2}) and reach numerical precision limit, or computational resources are exhausted, search failed.}
\end{center}

Candidate code must be then verified using symbolic methods, high-precision numerical confirmation
test, and ultimately proved using standard mathematical techniques. Above criterion do not
provide any numerical estimate for probability of successful identification.  However, we point out,
that even in case of possible misidentification, unexpected drop of error to machine epsilon
marks stop of the search anyway. This is because finding better approximation would require
formula with complexity already above threshold, given by intersection
of dotted lines in Fig.~\ref{Convergence2} at $N\simeq36$. Beyond that, number of formulas with identical
decimal expansion grows exponentially. This behavior mark search STOP criterion
for cases, where ''smoking gun'' feature from Fig.~\ref{SmokingGun} was not encountered. In practice
this still require a lot of computational resources, beyond capabilities of mid-range PC/laptop. That is why
my curves in Figs.~\ref{SmokingGun}-\ref{Convergence3} are still far from double epsilon, marked
with \texttt{DBL\_EPSILON} dotted line.

\begin{figure}
\includegraphics[width=0.69220062755534635386542199718279\textwidth]{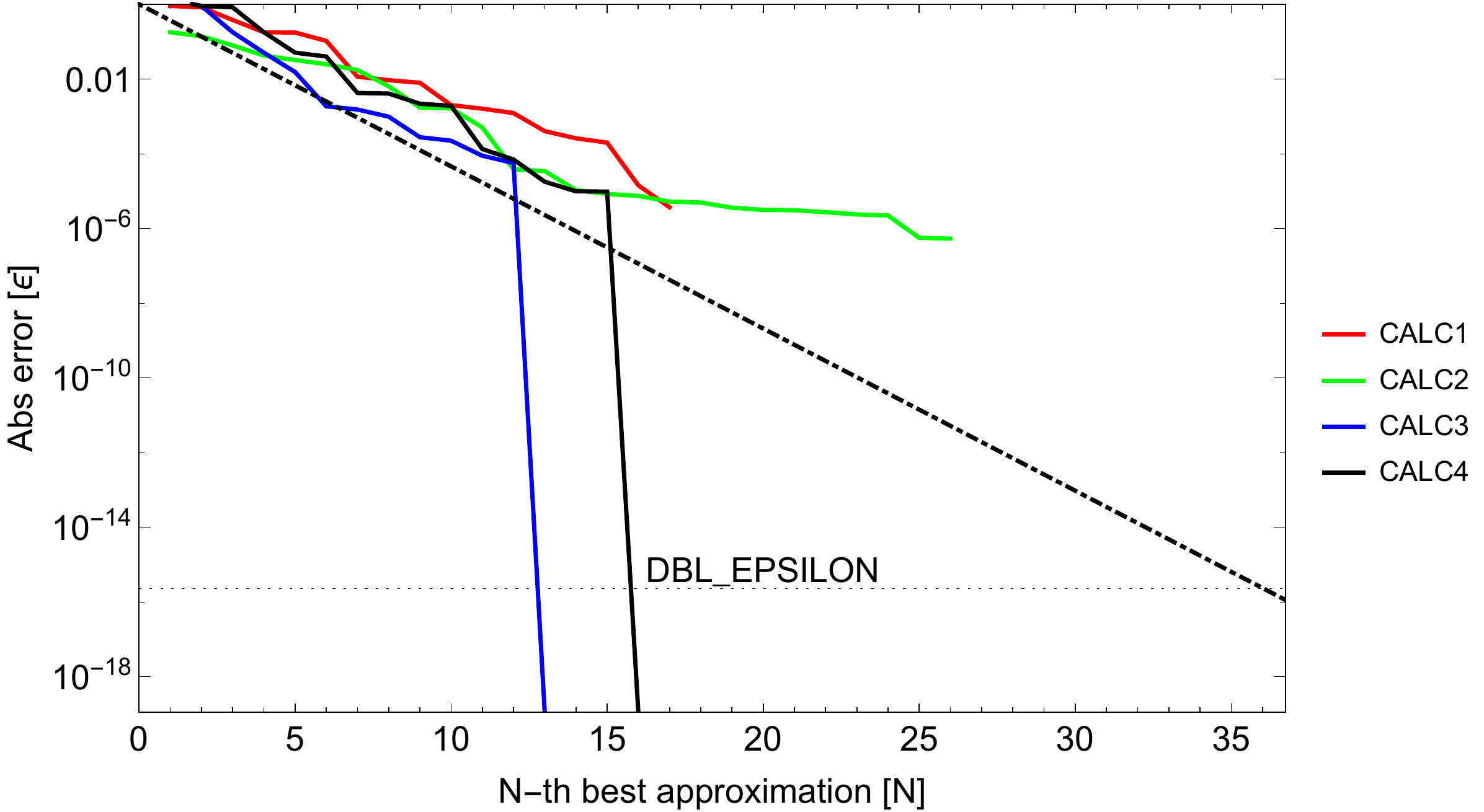}
\caption{\label{Convergence2} Similar to Fig.~\ref{SmokingGun}, but $K$ has been replaced
by number of subsequent best approximations $N$ found so far. This allows for direct comparison
of languages of different length. Possible identification is marked by error for next approximation significantly below $e^{-N}$.}
\end{figure}

Expected decrease of approximation error for next best one is $1/e \simeq 0.37$ of previous. Therefore, we  
use $e$-folding name for Criterion 1.

\begin{figure}
\includegraphics[width=\textwidth]{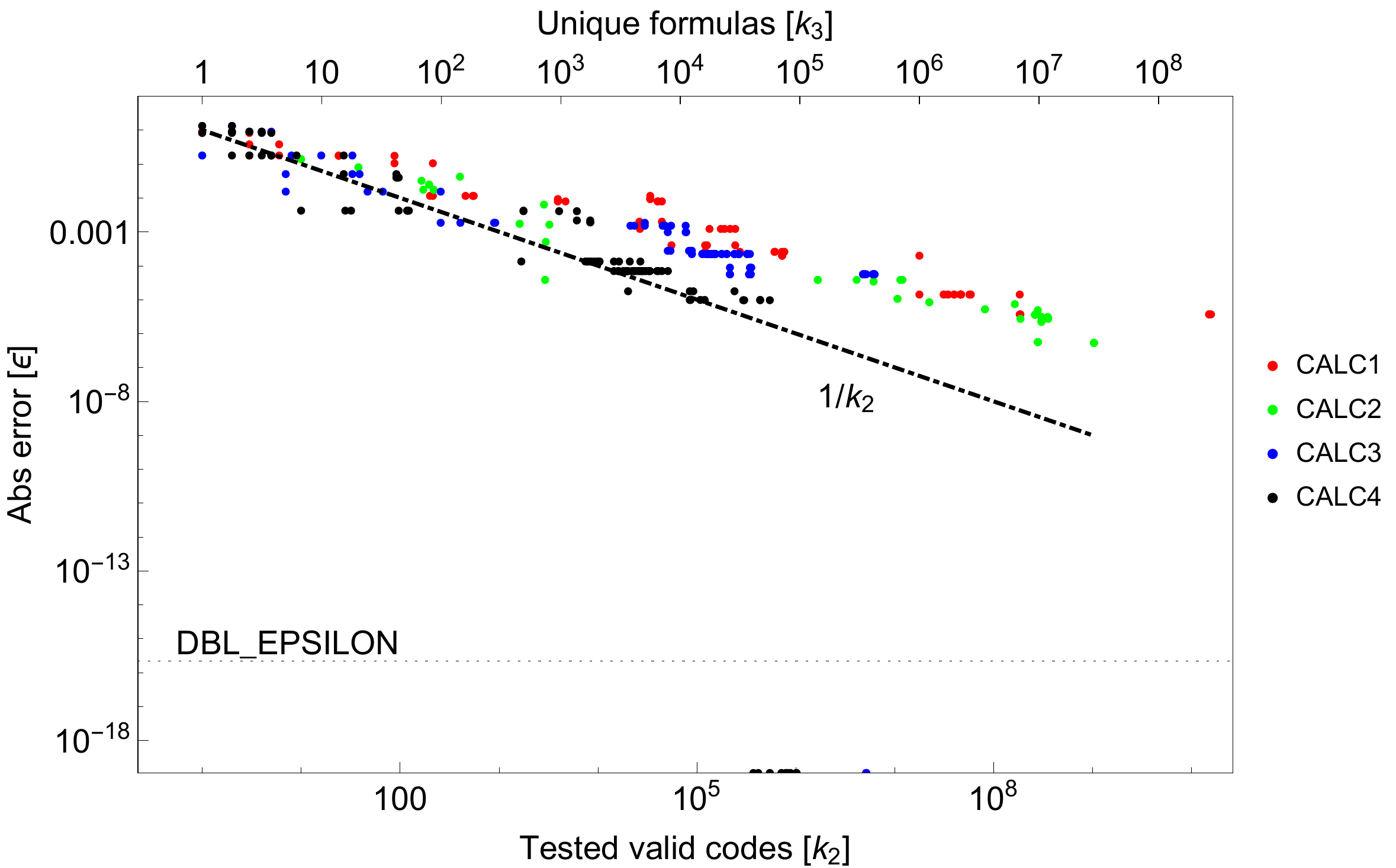}
\caption{\label{Convergence3} Similar to Fig.~\ref{Convergence2}, but $N$ has been replaced
by number $k_2$ of valid RPN codes tested so far. Power-law is observed, with $\epsilon \propto k_2^{-1}$. Upper
axis show estimated by power-law fit $k_3 \simeq k_2^{0.83}$ number of \textit{unique} formulas.}
\end{figure}

Remarkable observation is provided by yet another Figure~\ref{Convergence3}. Instead of $K$
or $N$ we used total number $k_2$ of \textit{valid} RPN codes tested before encountering
next approximation. Power law behavior is found, and band of data points (Fig.~\ref{Convergence3}) can be roughly approximated as proportional to $1/k_2$.  Therefore, obtaining definite negative answer for constant known with machine precision of $\sigma$ in terms of Criterion 1 require testing of $\sigma^{-1}$ codes. For double precision machine epsilon it is above $10^{15}$. You need either hundreds of CPU cores, or a lot of patience (days of search). Positive identification can be much faster,
of course, like for calculators 3, 4 in Figures~\ref{SmokingGun}, \ref{Convergence2}.

\end{section}

\begin{section}{Statistical properties of the exp-log numbers \label{Criterion2} }

Observation from Fig.~\ref{Convergence3} and results of previous section lead to another view of constant identification problem. It can be described as a random process. Consider following numerical \textit{Monte Carlo} experiment.
We generate pseudo-random numbers $\xi>0$ from exponential distribution with some scale $\lambda$
and Probability Distribution Function (thereafter PDF):
\begin{equation}
\label{Pexp}
P(\xi) = \frac{e^{-\xi/\lambda}}{\lambda}.
\end{equation}
Next, we compare $\xi$ with our target number $z>0$, 
and generate sequence of progressively better approximations. Surprisingly, observed behavior
is similar to Fig.~\ref{Convergence2}, but with larger fluctuations. 
Let's further assume $z$ is known with numerical precision $\sigma < z$.
In other words, true number is in the interval $[z-\sigma,z+\sigma]$. Assuming \eqref{Pexp}
we can easily obtain probability of random hit into vicinity $z$:
$$
P(z,\sigma, \lambda) = \int_{z-\sigma}^{z+\sigma} P(\xi) \; d \xi
$$
what gives average number of tries:
$$
k_2 = \frac{e^{z/ \lambda}}{2 \sinh{\sigma/\lambda}} \sim  \frac{const}{\sigma}.
$$
Above agrees qualitatively with result presented in Fig.~\ref{Convergence3} if $\sigma \ll z$.

However, statistical distribution of $\exp-\log$ numbers is unknown, PDF \eqref{Pexp} was chosen
intuitively. Statistical
properties of EL numbers should not depend on on language used to generate them, at least in the limiting case 
of large complexity. Numerical evidence, obtained by collecting real numbers generated
according to procedure presented in Sect.~\ref{enum}, is presented in Fig.~\ref{StatReals}. 
Numbers with $\operatorname{Im}(z) \neq 0$ were discarded\footnote{See Appendix~E in Supplementary Material for distribution of EL numbers on the complex plane.} to simplify analysis and presentation.  

\begin{figure}
\includegraphics[width=\textwidth]{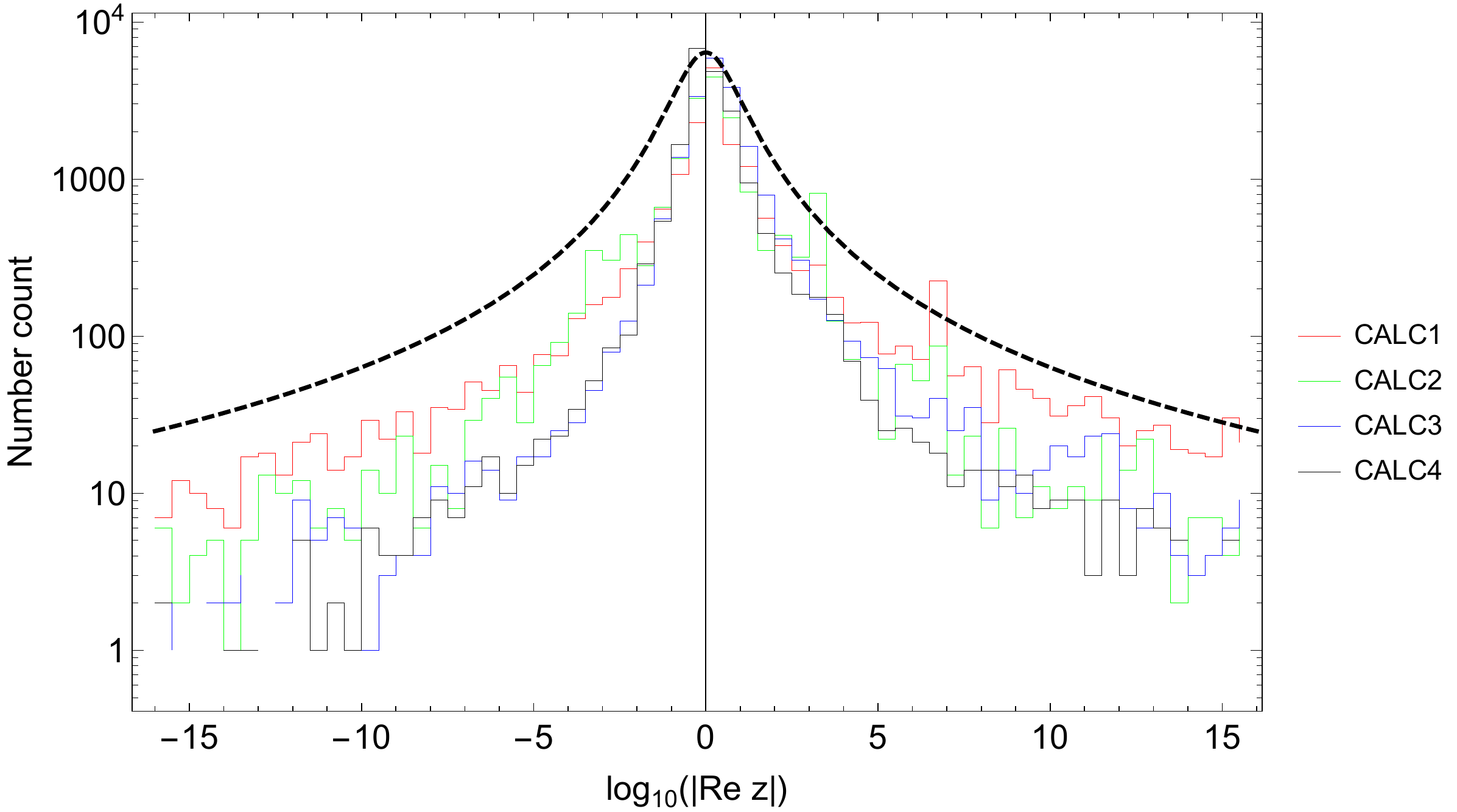}
\caption{\label{StatReals} Histogram of first 20k real numbers generated with use of
calculators 1-4. Dashed line show exponentially transformed Cauchy distribution \eqref{cauchy}.
} 
\end{figure}
    
None of the widespread distributions match numbers presented in Fig.~\ref{StatReals}. The most close
are: Pareto, Levy and Cauchy distributions. Noteworthy, all of
them have undefined mean and infinite dispersion. If we look at statistics of their base 10 logarithm,
it mimic t-Student, Cauchy and Laplace (double $\exp$) distributions. Probably this is indeed a brand new statistics, 
following some empirical distribution.  True distribution of real EL numbers (Fig.~\ref{StatReals}) is, however, quite well
described by transformed Cauchy distribution with PDF:
\begin{equation}
\label{cauchy}
P(x) = \frac{1}{\pi x} \frac{1}{1+(\ln{x})^2}.
\end{equation}
Therefore, later, statistical distribution of (real) EL numbers will be approximated
by Cauchy-like distribution \eqref{cauchy}.

Let us assume statistical distribution of real EL constants has PDF $P(x)$ empirically derived 
from histogram in Fig.~\ref{StatReals} or is given by \eqref{cauchy}. Let the target
number $z$ is know with precision $\sigma$, and statistical distribution of error for $z \pm \sigma$ is $Q(x, z, \sigma)$.
$Q$ might be normal or uniform distribution, for example. Then conditional probability $p_k$, that $k$-th tested formula \textbf{is not} random match for target value $z \pm \sigma$ is:
\begin{eqnarray}
p_k =  \prod_{i=1}^k   \left( 1-  \frac{\int_{-\infty}^{\infty} P(\xi) Q(\xi) \; d \xi}{Q(z)} \right) = \\ \nonumber
=
\left( 1-  \frac{\int_{-\infty}^{\infty} P(\xi) Q(\xi) \; d \xi}{Q(z)} \right)^k.
\end{eqnarray}

Then, replacing $P(\xi) \simeq P(z)$ and integrating,  likelihood $\mathcal{L}$ for proper identification
of target constant $z$ with i-th unique value $x_i$ is:
\begin{equation}
\label{Id_likelihood}
\mathcal{L} = \left[ 1-   \frac{P(z)}{Q(z)} \right]^{k_3} Q(x_i, z, \sigma),
\end{equation}
where $i=k_3$, i.e. number of unique values tested so far. To understand \eqref{Id_likelihood},
it might be approximated by
\begin{equation}
\label{Id_likelihood2}
\left( 1- \kappa \, \sigma \; P(z) \; k_3 \right) Q(x_i, z, \sigma),
\end{equation}
assuming $\sigma \ll z$ and using Maclaurin expansion $(1+\alpha)^k \simeq 1 + \alpha k + \ldots$. Value of $\kappa$ depends on error distribution $Q$, e.g.,  $\kappa=2$ for uniform and $\kappa=\sqrt{2 \pi} \simeq 2.5$ 
for normal distribution.

Likelihood \eqref{Id_likelihood2} becomes zero for large $k_3$, where search for $k_3>1/\sigma$ is pointless anyway due to Criterion 1 (Sect.~\ref{Criterion1}). The most intriguing
application of \eqref{Id_likelihood} is when $\sigma$ is still large compared with machine epsilon,
and has uncertainty of statistical nature. This allow for selection of maximum likelihood formula(s) in marginal sense,
i.e. for $k_3 \sim \sigma^{-1}$. Because calculation of $k_3$ requires storage of all previously computed values,
it is reasonable to replace it with $k_2$ using power-law fit from Fig.~\ref{k2k3}:

\begin{equation}
\label{Id_likelihood3}
\left[  1- \kappa \;  {k_2}^p \;  \sigma\;  P(z) \right] \times  Q(x_i, z, \sigma).
\end{equation}

Noteworthy, \eqref{Id_likelihood3} as a function of $k_2$, i.e. number of tested
syntactically correct RPN codes, must have a maximum. Formula for $x_i$ with maximum probability is well defined. For limited precision maximum is very pronounced (Fig.~\ref{likelihood}), but for very small $\sigma$ we might not be able to reach it at all, due to limited computational resources (Fig.~\ref{likelihood}, solid blue). Equations (\ref{Id_likelihood}-\ref{Id_likelihood3}) can be understood as follows. If we find remarkably simple
and elegant formula at the very beginning of the search, which is however quite far from target $z$, e.g.,
several $\sigma$'s in Gaussian distribution, we reject it on the basis of improbable error in measurement/calculation. Similarly, if we find formula well within error range, e.g. $4 \sigma$, but after search covering billions of formulas, we reject it expecting random coincidence. Somewhere in the middle lays
optimal formula, with maximum likelihood, estimated with use of \eqref{Id_likelihood}. In practice,
likelihood is very small, and $\log$-likelihood is more convenient: 
$$
\log \mathcal{L} = -k_3 \ln \left(1 -  \frac{P(z)}{Q(z,z,\sigma)} \right) + \ln Q(x_i,z,\sigma).
$$
Expanding $\ln{(1-\epsilon)} \simeq -\epsilon$ and assuming error distribution $Q$ is Gaussian, I have
derived following useful approximation for $\log$-likelihood:
\begin{equation}
\label{logL}
\log \mathcal{L} = k_3 \frac{P(z)}{Q(z,z,\sigma)} - \ln{(\sqrt{2 \pi} \sigma)} - \frac{(x_i-z)^2}{2 \sigma^2}.
\end{equation}

We note, that probability of the identification related to \eqref{Id_likelihood} could be, in principle, validated
directly, especially for single precision floats, as there are only $2^{32}$ of them. Using floats
as bins for sequentially generated EL numbers, we can fill them with numbers associated with their
complexity.

\begin{figure}
\includegraphics[width=0.5\textwidth]{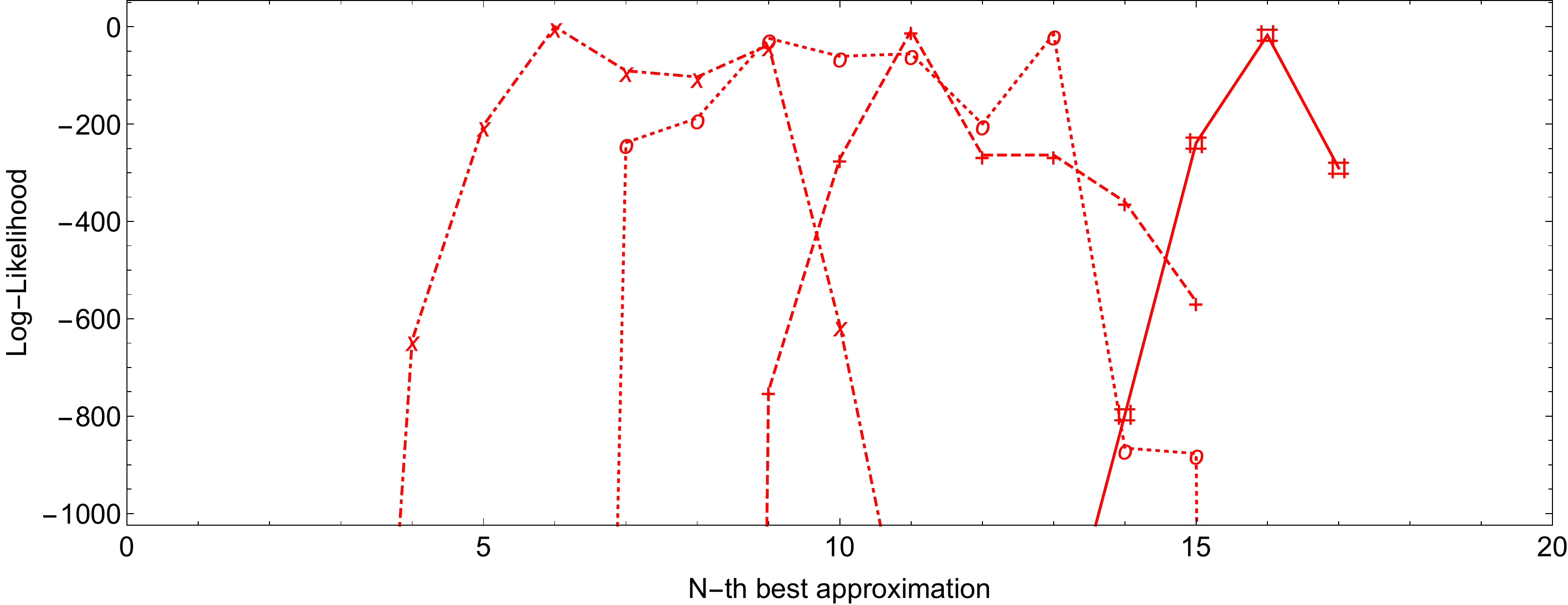}~\includegraphics[width=0.5\textwidth]{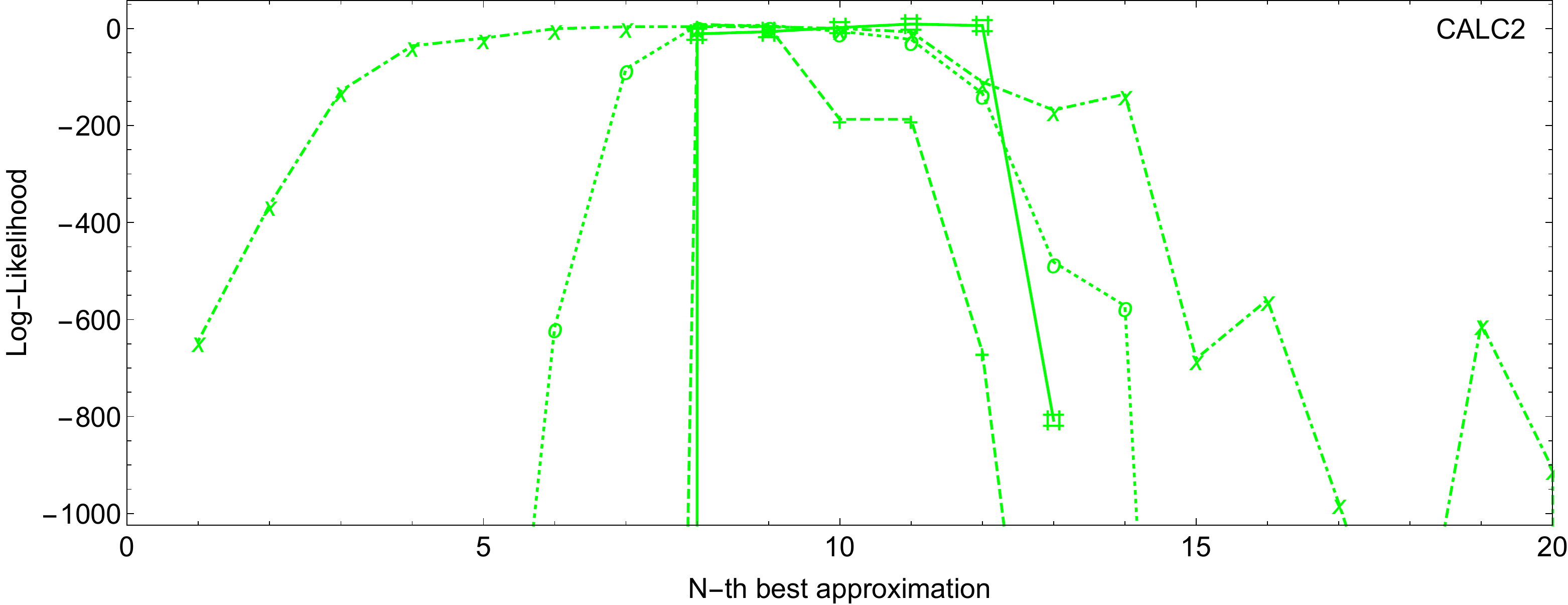}\\
\includegraphics[width=0.5\textwidth]{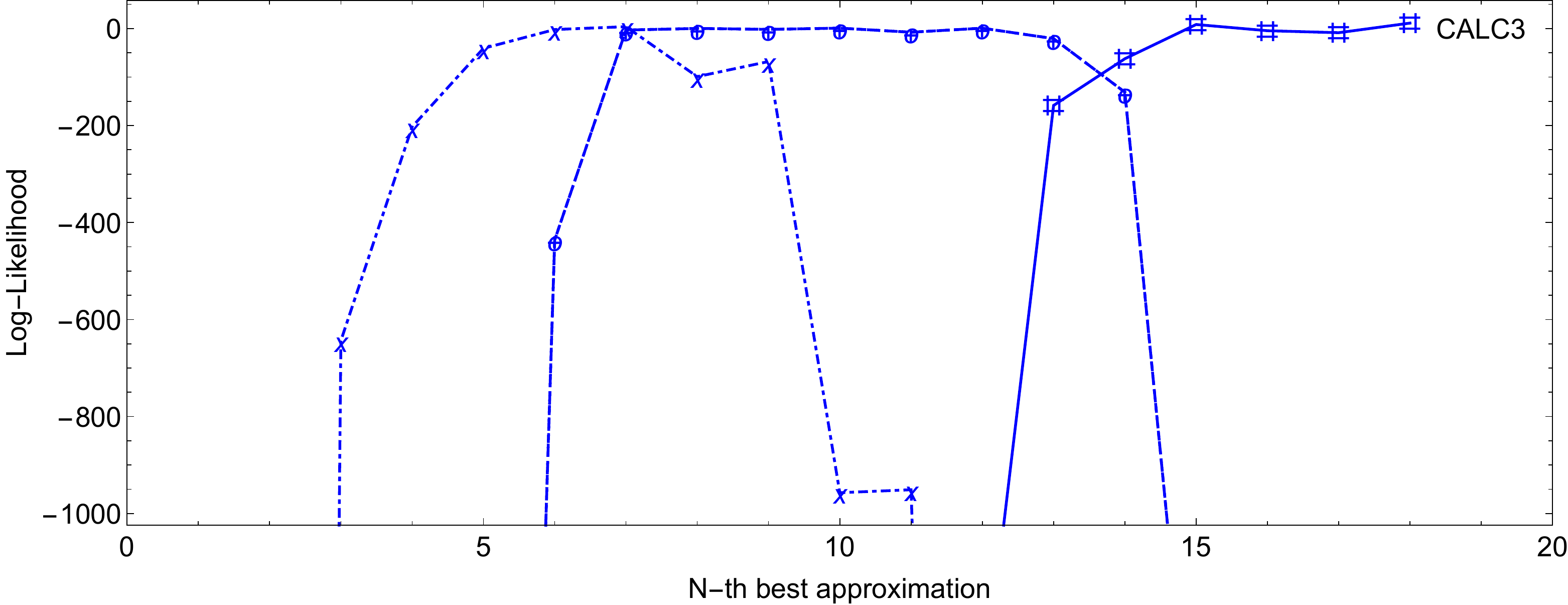}~\includegraphics[width=0.5\textwidth]{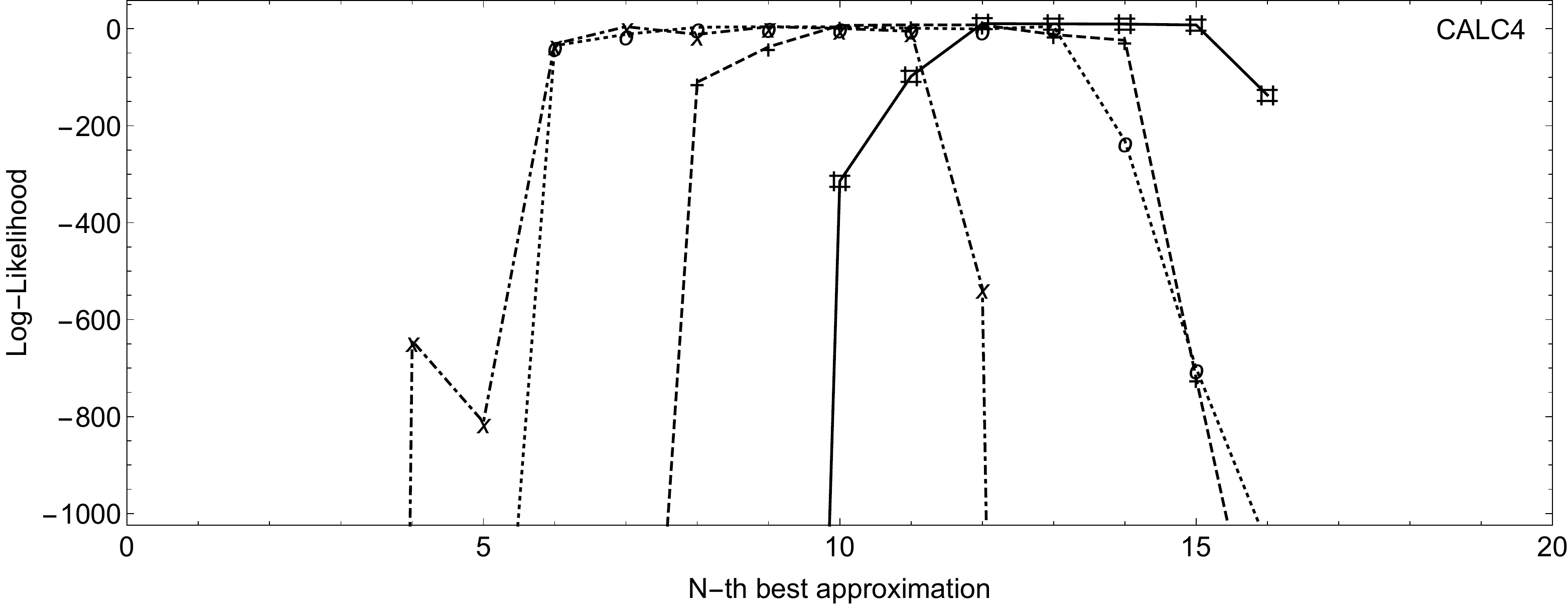}
\caption{
\label{likelihood} 
Log-likelihood of subsequent best $\exp-\log$ approximations, as estimated from \eqref{logL}.
Red, green, blue and black lines are for calculators 1,2,3 and 4, respectively. Target values are approximations
to our introduction example $z=\sqrt{2}^{\sqrt{3}}$ in form $z \pm \sigma$ with: $z=1.82 \pm 0.005$ (dot-dashed, x), $z=1.823 \pm 0.0005$ (dotted, o), $z=1.8226 \pm 0.00005$ (dashed, +), $z=1.82263 \pm 0.000005$ (solid, \#).
}
\end{figure}

Discussion above provides second criterion for constant identification:
\begin{center}
Criterion 2:\\ 

Identification candidate:
\textit{if likelihood given by \eqref{Id_likelihood} has reached maximum, formula has the highest probability,
and should be returned.}

\end{center}

One might return a few highest likelihood formulas near maximum as well, or one with largest value so far, if computational
resources are limited. Noteworthy, likelihood value provide also relative quantitative estimate of identification probability. 
We may estimate likelihood \eqref{logL}, using calculator \eqref{CALC3}, for $z=1.82263\pm0.00005$ (Fig.~\ref{likelihood}, solid blue line) to be identified as $\sqrt{2}^{\sqrt{3}}$
$$\log \mathcal{L} = 8.33 \qquad   (\mathcal{L} = 4 \times 10^{3})$$ 
compared to 
$$\log \mathcal{L} = -62.3 \qquad  (\mathcal{L} = 9 \times 10^{-28})$$
for $(\ln{4})^{\ln{2\pi}}$. From Criterion 2, the former is more than $10^{30}$ times more probable than the latter.
Without this sort of ''ranking'', discussion, if e.g. $\sinh ^{-1}\left(\phi  \sinh ^{-1}(\pi )\right)$
is ''simpler'' or ''more elegant'' compared to, e.g., $\sin( \cosh{4} ) + \sqrt{\tanh{5}}$ might continue indefinitely,
leading to nowhere. Likelihood \eqref{likelihood} or \eqref{logL} provide numerical form of the Occam's razor. It can be
applied automatically, within software (e.g. Computer Algebra System) environment.

\end{section}

\begin{section}{Constant recognition as data compression \label{Criterion3} }

Sequence of progressively better approximations in form of RPN calculator
codes can be viewed as a form of lossy compression. If exact formula
is found, then compression becomes lossless. In the intermediate case,
compression ratio provides measure, how good is some formula to recover
decimal expansion. This process is illustrated in Table~\ref{compress}. Calculator
3 defined in eq.~\eqref{CALC3} has been used, because both decimal expansion of the target number
$z$ and RPN code are strings of the same base-10 digits, i.e. \texttt{0123456789}. Therefore,
compression ratio is simply a number of correct digits divided by code length $K$.

\begin{table}
\begin{tabular}{rrcc}
Numerical & Code & Compression          & Formula\\
value     &      & ratio                &   \\[2mm]
\hline \\[2mm]
         3.141592653589793238512 &	 0 & 0.00       & $\pi$\\
         2.718281828459045235428 &    1 & 0.00      & $e$\\
         0.000000000000000000000 &    2 & 0.00      & $0$\\
         2.000000000000000000000 &    7 & 0.00      & $2$\\
\textbf{1}.718281828459045235428 &  164 & 0.33      & $e-1$\\
\textbf{1}.772453850905516027310 &  809 & 0.33      & $\sqrt{\pi}$\\
\textbf{1}.648721270700128146893 &  819 & 0.33      & $\sqrt{e}$\\
\textbf{1.8}37877066409345483606 & 0043 & 0.50      & $\ln{(\pi+\pi)}$\\ 
\textbf{1.8}37877066409345483606 & 7053 & 0.50      & $\ln{(2 \pi)}$\\
\textbf{1.82}0796326794896619256 &08485 & 0.60      & $\frac{1}{2}(\pi+\frac{1}{2})$\\ 
\textbf{1.82}0796326794896619256 &80485 & 0.60      & \\ 
\textbf{1.82}0796326794896619256 &80845 & 0.60      & \\ 
\textbf{1.82}0796326794896619256 &88045 & 0.60      & \\ 
\textbf{1.82}1126701185962651818 & 0338975 & 0.43   & $2^{1-\ln{\ln{\pi}}}$\\
\textbf{1.82}1126701185962651818 & 7033895 & 0.43   & \\
\textbf{1.82}1126701185962651818 & 8303975 & 0.43   & $2 (\ln{\pi})^{-\ln{2}}$\\
\textbf{1.82}4360635350064073446 & 8819745 & 0.43   & \\
\textbf{1.82}4360635350064073446 & 8781945 & 0.43   & \\
\textbf{1.82}4360635350064073446 & 8197485 & 0.43   & \\
\textbf{1.82}4360635350064073446 & 7819485 & 0.43   & \\
\textbf{1.82}1126701185962651818 & 7830395 & 0.43   & \\
\textbf{1.82}1662858741926632288 & 6091579 & 0.43   & \\
\textbf{1.822}361069544464599575 & 2298979 & 0.57   & \\
\textbf{1.822}413909696397869321 & 77408934 & 0.50  & \\
\textbf{1.822}722133555469366033 & 80790539 & 0.50  & \\
\textbf{1.8226}90334737686312645 & 004377539 & 0.56 & \\
\textbf{1.822634654966242214}488 & 888854979 & 2.11 & 
\end{tabular}
\caption{\label{compress}} 
\end{table}

In Table~\ref{compress}, compression ratio of the target number $z=\sqrt{2}^{\sqrt{3}} \simeq 1.8226346549662422143937682155941\ldots$
in form of $n=10$ RPN calculator codes \eqref{CALC3} is shown.

Buttons/operations were assigned as follows: \\
0 $\to$ \pmb{Pi}, \\
1 $\to$ \pmb{E}, \\
2 $\to$ \pmb{I}, \\
3 $\to$ \pmb{Log},\\ 
4 $\to$ \pmb{Plus}, \\ 
5 $\to$ \pmb{Times}, \\ 
6 $\to$ \pmb{-1}, \\
7 $\to$ \pmb{2}, \\
8 $\to$ \pmb{1/2},\\ 
9 $\to$ \pmb{Power}.

In fact, last code in Table~\ref{compress} \texttt{888854979}, with pronounced compression ratio of $19/9\simeq2.11 > 1$, is equivalent
to exact formula, with RPN sequence \textonehalf, \textonehalf, \textonehalf, \textonehalf, $\times$, $+$, \pmb{Power}, 2, \pmb{Power},
i.e., $2^{\sqrt{3}/2}$.

In general case, compresion ratio $r$ is given by:
\begin{equation}
\label{r}
r = \frac{-\log_{10} \max{(\epsilon, \sigma)}}{K \; \log_{10} n},
\end{equation}
where $\epsilon$ is absolute precision of the approximation, $K$ - RPN code length, $n$ - number of calculator buttons.

Now I can formulate the third criterion for constant recognition:
\begin{center}
Criterion 3:\\ 

Identification candidate:
\textit{if compression ratio given by \eqref{r} is $r \gg 1$, or $r$ reaches maximum in the course of search, then formula 
should be returned.}\\

Failure: \textit{if $r \ll 1$ formula/code is unlikely match.}

\end{center}

Criterion 3 has an advantage of being very simple. It do not require recorded history of search, like Criterion 1 (Sect.~\ref{Criterion1}),
of searching for maximum and knowledge of statistics, like Criterion 2 (Sect.~\ref{Criterion2}). If indeed $r$ has a maximum, then it strenghten identification, but this is not required. You might ask for compresion ratio of any combination of decimal expansion and formula. The only required action is to compile
formula to RPN code \eqref{CALC} or similar one. Therefore, it will work with any searching method, e.g: \textit{Monte~Carlo}, genetic or shortest path tree algorithms. It is weak compared to $e$-folding or statistical criteria, but could 
easily exclude most of formulae produced in variety of software, especially very complicated ones, and those including large ($i \gg 2$) integers.

\end{section}

\begin{section}{Blind test example for three criteria}

Three criteria proposed in the article provide robust tool for decimal
constant identification. 

\begin{figure}
\includegraphics[width=\textwidth]{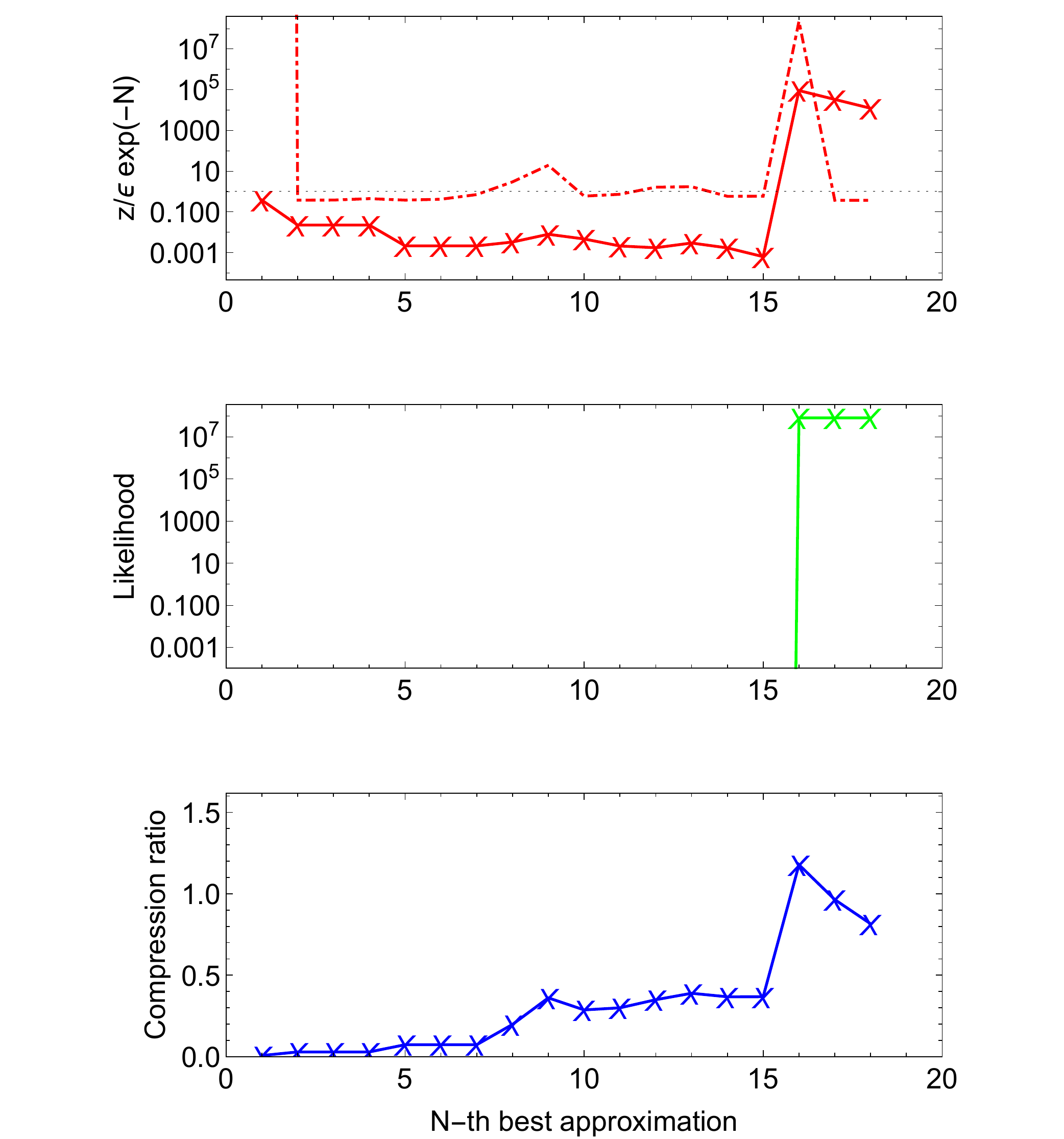}
\caption{\label{3C} Recognition indicators as a function of $N$. Solid red line shows absolute $e$-folding $e_1$,
while dotted red line relative $e$-folding $e_2$ (upper panel). Green line show likelihood \eqref{logL} (middle panel),
and blue compression ratio \eqref{r} (bottom panel).}
\end{figure}

Process of recognition, applying all three criteria,  is illustrated in Fig.~\ref{3C}, using blind-test
target value of $z=201.06192983$. Assuming all digits are correct, we adopted (gaussian) error $\sigma=0.000000005$.
Using calculator \eqref{CALC3}, we plot three indicators of matching quality. For $e$-folding measure we use
$$
e_1 = \frac{z}{\epsilon_N} e^{-N},
$$
and additionally:
$$
e_2 = \frac{\epsilon_{N-1}}{\epsilon_N} \frac{1}{e},
$$
where $N$-th best approximation error is $\epsilon_N$.
For likelihood, we used \eqref{logL}, and for compression ratio \eqref{r}. 
Clearly, three subsequent data points, for $N=16,17,18$, stand out (Fig.~\ref{3C}).

From first criterion perspective (Sect.~\ref{Criterion1}), error dropped nearly $10^5$ times compared to expected $e^{-N}$
(Fig.~\ref{3C}, upper panel, solid red).
Is also $10^8$ times smaller compared to previously found $x_{15}=e^e (1/2+\pi)^2$, while statistically anticipated decrease was $1/e$ (Fig.~\ref{3C}, upper panel, dotted red). Using
criterion 2 (Sect.~\ref{Criterion2}) we found likelihood many many orders of magnitude larger compared
to any other formulas. Three data points with nearly equal $\log L\sim 18.193$ (Fig.~\ref{3C}, middle panel, green crosses),
16,17,18-th best approximations, are in fact the same formula typed using different RPN sequences. Differences
are due to round-off errors. Shortest one
is preferred in terms of criterion 3 (Sect.~\ref{Criterion3}), see Fig.~\ref{3C}, lower panel. Constant $z$ is then unambiguously recognized by all three criteria as $64 \pi$, with no other candidates in sight.

\end{section}

\begin{section}{Main results and conclusions}

Article is devoted to long-standing problem with recognition of constant resulting from some mathematical
formula given only its decimal expansion. Major results presented in the article are
\begin{enumerate}
\item well-posed formulation of the recognition task in the form of reversing calculator
button sequence
\item discovery of the minimal button set \eqref{CALC1}, which must be retained in any
kind of general symbolic regression software
\item three criteria selecting the most probable candidate formula.
\end{enumerate}

Blind test from previous section show the way to precise formulation of decimal constant recognition problem, as a
task reverse to pushing button sequence, and its solution. Using three proposed criteria we are 
able to judge which formulas, matching given floating-point constant, are the most likely. In original
formulation they require enumeration of all possible codes with growing complexity, but statistical 
and compression criteria are in fact independent of the method used to obtain expression. They require
formula to be compiled into RPN code. Then, Criterion 3 \eqref{r} can be applied directly, and Criterion 2 \eqref{logL} indirectly, using code length $K$ to find upper limit on $k_1,k_2,k_3$ (Fig.~\ref{k2k3}).

Calculator used to enumerate codes can be arbitrary, as long as it includes ''irreducible'' buttons equivalent to \eqref{base-3}
or \eqref{base-4}. However, search results (required depth in particular) depend on calculator definition. This is especially visible in search of  human-provided test cases, with strong bias towards decimal numerals, and repulsive reaction of human brain to power-towers and nested (exponential) function compositions. Unfortunately, the latter are strong approximators, with multi-layer neutral networks being notable example using sigmoidal functions \cite{backpropagation}. Therefore, for numbers provided by humans, full calculator \eqref{CALC4} is usually the best, for mathematical/physical formulas \eqref{CALC3} and for truly random expressions \eqref{CALC1} or \eqref{CALC2}.

As I mentioned in the Introduction,  while searching for constants, we are in fact traversing all elementary functions of one complex variable. This is especially  visible in the definition of calculator \eqref{base-3}, where $x$ is arbitrary. Therefore, 
mathematical software developed to solve constant identification problem, could be immediately
applied to identification of functions of one variable \cite{SymbolicRegressionPackage}. Extending our calculators 1-4 with additional
,,buttons'' representing input variables and model parameters/weights is also quite easy. This is a task for further research, with many more
potential applications.  However, due to increased dimensionality, convergence analysis and identification criteria are not as easy to handle
as for numerical constant case, while simultaneously use of brute-force enumeration search becomes inefficient. Future research on identification/approximation of general functions should therefore concentrate on genetic or Monte Carlo algorithms, based on presented ,,calculator'' approach at its core.

\end{section}

\bibliographystyle{plain}
\bibliography{ConstantRecognition}

\begin{thebibliography}{32}

\bibitem{backpropagation} Terrence J. Sejnowski. 2018. The Deep Learning Revolution. The MIT Press.

\bibitem{ADAM} Kingma, D. P. and Ba, J., Adam: A Method for Stochastic Optimization, arXiv e-prints arXiv:1412.6980v9 [cs.LG], 2014.

\bibitem{AIFeynmann} Silviu-Marian Udrescu and Max Tegmark, AI Feynman: A physics-inspired method for symbolic regression, Science Advances  15 Apr 2020, Vol. 6, no. 16, eaay2631, DOI: 10.1126/sciadv.aay2631

\bibitem{Eurega} Michael Schmidt and Hod Lipson. Distilling free-form natural laws from experimental data. Science, 324(5923):81–85, 2009

\bibitem{SymbolicRegressionPackage} A. Odrzywolek, Symbolic Regression Package for Mathematica, (2021), https://github.com/VA00/SymbolicRegressionPackage

\bibitem{RamanujanMachine} Raayoni, G., Gottlieb, S., Manor, Y. et al. Generating conjectures on fundamental constants with the Ramanujan Machine. Nature 590, 67-73 (2021). https://doi.org/10.1038/s41586-021-03229-4

\bibitem{BorweinBailey} David~H. {Bailey} and {Borwein}~Jonathan M. \newblock {Exploratory Experimentation and Computation}. \newblock {\em Notices of the AMS}, 58:1410--1419, November 2011.

\bibitem{plouffe}  David H. Bailey and Simon Plouffe, Recognizing Numerical Constants, http://www.cecm.sfu.ca/organics/papers/bailey/paper/html/paper.html

\bibitem{Maple} Maple (2017.3). Maplesoft, a division of Waterloo Maple Inc., Waterloo, Ontario.

\bibitem{Mathematica} Wolfram Research, Inc., Mathematica, 12.3.0 for Microsoft Windows (64-bit) (May 10, 2021), Champaign, IL (2021).

\bibitem{symPy} Meurer A, Smith CP, Paprocki M, Čertík O, Kirpichev SB, Rocklin M, Kumar A, Ivanov S, Moore JK, Singh S, Rathnayake T, Vig S, Granger BE, Muller RP, Bonazzi F, Gupta H, Vats S, Johansson F, Pedregosa F, Curry MJ, Terrel AR, Roučka Š, Saboo A, Fernando I, Kulal S, Cimrman R, Scopatz A. (2017) SymPy: symbolic computing in Python. PeerJ Computer Science 3:e103

\bibitem{RIES} Robert Munafo, RIES - find algebraic equations, given their solution. https://mrob.com/pub/ries/index.html Accessed: 2021-06-08.

\bibitem{ISCwayback} Inverse Symbolic Calculator, http://wayback.cecm.sfu.ca/projects/ISC/ISCmain.html

\bibitem{euler} Bruce J. Petrie,
Leonhard Euler’s use and understanding of mathematical transcendence,
Historia Mathematica, Volume 39, Issue 3, 2012, Pages 280-291, https://doi.org/10.1016/j.hm.2012.06.003

\bibitem{double} D. Hough, "The IEEE Standard 754: One for the History Books" in Computer, vol. 52, no. 12, pp. 109-112, 2019. doi: 10.1109/MC.2019.2926614
\bibitem{complexity} Ray J. Solomonof, INFORMATION AND CONTROL. Volume 7, No. 2, June
1964, pp. 224–254. Copyright by Academic Press Inc.
\bibitem{closed} J. M. Borwein and R. E. Crandall. Closed forms: What they are and why we care. Notices of the American Mathematical Society, 60:50–65, 2013.
\bibitem{chow} Timothy Y. Chow. What is a closed-form number? The American Mathematical Monthly, 106(5):440–448, 1999.

\bibitem{brute} Jonathan Schaeffer, H.Jaap van den Herik,
Games, computers, and artificial intelligence,
Artificial Intelligence,Volume 134, Issues 1–2,2002,Pages 1-7,ISSN 0004-3702,
https://doi.org/10.1016/S0004-3702(01)00165-5.

\bibitem{zaremba} Wojciech Zaremba, Karol Kurach, and Rob Fergus. 2014. Learning to discover efficient mathematical identities. In Proceedings of the 27th International Conference on Neural Information Processing Systems - Volume 1 (NIPS’14). MIT Press, Cambridge, MA, USA, 1278–1286.
\bibitem{BorweinBailey2} David H. Bailey, Jonathan M. Borwein, Alexander D. Kaiser,
Automated simplification of large symbolic expressions,
Journal of Symbolic Computation, Volume 60, 2014, Pages 120-136, https://doi.org/10.1016/j.jsc.2013.09.001
\bibitem{combinatorial} A. Dey, J. Stenberg, P. Dandekar, R. Jain,
\newblock {A combinatorial study of experimental analysis and mathematical modeling: How do chitosan nanoparticles deliver therapeutics into cells?}, Carbohydrate Polymers, Volume 229, 2020,115437, https://doi.org/10.1016/j.carbpol.2019.115437
\bibitem{bottou} L. Bottou. From machine learning to machine reasoning. Machine Learning, 94(2):133-149, 2014. 
\bibitem{epyc} AMD EPYC™ 7002 Series Processors and DELL POWEREDGE™ R6525 Servers Set World Record on Industry Standard Decision Support Benchmark https://www.amd.com/system/files/documents/tpc-h-dell-3tb-exasol.pdf
\bibitem{gpu} Ben van Werkhoven,
Kernel Tuner: A search-optimizing GPU code auto-tuner,
Future Generation Computer Systems,
Volume 90,
2019,
Pages 347-358,
ISSN 0167-739X,
https://doi.org/10.1016/j.future.2018.08.004.	
\bibitem{simd1} Hiroshi Watanabe, Koh M. Nakagawa,
SIMD vectorization for the Lennard-Jones potential with AVX2 and AVX-512 instructions,
Computer Physics Communications,
Volume 237,
2019,
Pages 1-7,
ISSN 0010-4655,
https://doi.org/10.1016/j.cpc.2018.10.028.
\bibitem{simd2} Hossein Amiri, Asadollah Shahbahrami,
SIMD programming using Intel vector extensions,
Journal of Parallel and Distributed Computing,
Volume 135,
2020,
Pages 83-100,
ISSN 0743-7315,
https://doi.org/10.1016/j.jpdc.2019.09.012.
\bibitem{fpga1} Övünç Polat, Tülay Yıldırım,
FPGA implementation of a General Regression Neural Network: An embedded pattern classification system,
Digital Signal Processing,
Volume 20, Issue 3,
2010,
Pages 881-886,
ISSN 1051-2004,
https://doi.org/10.1016/j.dsp.2009.10.013.
\bibitem{fpga2} Peter Irgens, Curtis Bader, Theresa Lé, Devansh Saxena, Cristinel Ababei,
An efficient and cost effective FPGA based implementation of the Viola-Jones face detection algorithm,
HardwareX,
Volume 1,
2017,
Pages 68-75,
ISSN 2468-0672,
https://doi.org/10.1016/j.ohx.2017.03.002.
\bibitem{fpga3} Omair Inam, Abdul Basit, Mahmood Qureshi, Hammad Omer,
FPGA-based hardware accelerator for SENSE (a parallel MR image reconstruction method),
Computers in Biology and Medicine,
Volume 117,
2020,
103598,
ISSN 0010-4825,
https://doi.org/10.1016/j.compbiomed.2019.103598.
\bibitem{quant1} Pranav Santosh Menon, M. Ritwik,
A Comprehensive but not Complicated Survey on Quantum Computing,
IERI Procedia,
Volume 10,
2014,
Pages 144-152,
ISSN 2212-6678,
https://doi.org/10.1016/j.ieri.2014.09.069.
\bibitem{quant2}  Sándor Imre,
Quantum computing and communications – Introduction and challenges,
Computers \& Electrical Engineering,
Volume 40, Issue 1,
2014,
Pages 134-141,
ISSN 0045-7906,
https://doi.org/10.1016/j.compeleceng.2013.10.008.
\bibitem{quant3}  Hans De Raedt, Fengping Jin, Dennis Willsch, Madita Willsch, Naoki Yoshioka, Nobuyasu Ito, Shengjun Yuan, Kristel Michielsen,
Massively parallel quantum computer simulator, eleven years later,
Computer Physics Communications,
Volume 237,
2019,
Pages 47-61,
ISSN 0010-4655,
https://doi.org/10.1016/j.cpc.2018.11.005.
\bibitem{ai1} Marta Garnelo, Murray Shanahan,
Reconciling deep learning with symbolic artificial intelligence: representing objects and relations,
Current Opinion in Behavioral Sciences,
Volume 29,
2019,
Pages 17-23,
ISSN 2352-1546,
https://doi.org/10.1016/j.cobeha.2018.12.010.
\bibitem{ai2} Ernest Davis,
Ethical guidelines for a superintelligence,
Artificial Intelligence,
Volume 220,
2015,
Pages 121-124,
ISSN 0004-3702,
https://doi.org/10.1016/j.artint.2014.12.003.
\bibitem{ai3} Ben Goertzel, Artificial Intelligence 171 (2007) 1161–1173
\bibitem{intelligence} José Hernández-Orallo, David L. Dowe,
Measuring universal intelligence: Towards an anytime intelligence test,
Artificial Intelligence,
Volume 174, Issue 18,
2010,
Pages 1508-1539,
https://doi.org/10.1016/j.artint.2010.09.006.
\bibitem{SR} Sohrab Towfighi, Symbolic regression by uniform random global search, Neural and Evolutionary Computing (cs.NE) Cite as:	arXiv:1906.07848 [cs.NE]
\bibitem{piday} Ancient pi calculator gets a modern twist for pi day,
New Scientist, Volume 217, Issue 2908, 2013, Page 4, ISSN 0262-4079,
https://doi.org/10.1016/S0262-4079(13)60645-4.



\bibitem{glibc} Sandra Loosemore with Richard M. Stallman, Roland McGrath, Andrew Oram, and Ulrich Drepper, The GNU C Library Reference Manual for version 2.31, https://www.gnu.org/software/libc/manual/
\bibitem{Knuth} Donald Knuth, The Art of Computer Programming, Volume 4, Fascicle 2: Generating All Tuples and Permutations,  Addison-Wesley Professional; 1 edition (February 24, 2005), ISBN-13: 978-0201853933, ISBN-10: 0201853930
\bibitem{Tarski} Alfred Tarski. A decision method for elementary algebra and geometry. U. S. Air Force Project Rand, R-109. Prepared for publication by J. C. C. McKinsey. Litho-printed. The Rand Corporation, Santa Monica, California, 1948
\bibitem{e} Eli Maor, The story of a number $e$, Princeton University Press, 1994, ISBN 0-691-03390-0
\bibitem{const} Eric W. Weisstein, Constant, From MathWorld--A Wolfram Web Resource. http://mathworld.wolfram.com/Constant.html
\bibitem{briggs} Denis Roegel. A reconstruction of the tables of Briggs Arithmetica logarithmica (1624). [Research Report] 2010. ffinria-00543939f
\bibitem{RPNorig} Jan Lukasiewicz (1957). Aristotle's Syllogistic from the Standpoint of Modern Formal Logic. Oxford University Press. (Reprinted by Garland Publishing in 1987. ISBN 0-8240-6924-2)
\bibitem{RPN} C. L. Hamblin, Translation to and from Polish Notation, The Computer Journal, Volume 5, Issue 3, November 1962, Pages 210–213, https://doi.org/10.1093/comjnl/5.3.210
\bibitem{tree} L. Tychonievich L (2013) Enumerating Trees. URL
https://www.cs.virginia.edu/~lat7h/blog/posts/434.html
\bibitem{richardson} Richardson, Daniel (1968). "Some undecidable problems involving elementary functions of a real variable". Journal of Symbolic Logic. 33 (4). Association for Symbolic Logic. pp. 514-520. doi:10.2307/2271358. JSTOR 2271358.
\bibitem{OMP} Leonardo Dagum and Ramesh Menon. 1998. OpenMP: An Industry-Standard API for Shared-Memory Programming. IEEE Comput. Sci. Eng. 5, 1 (January 1998), 46-55. DOI:https://doi.org/10.1109/99.660313
\bibitem{learning} B. Jonsson, M. Norqvist, Y. Liljekvist, J. Lithner,
Learning mathematics through algorithmic and creative reasoning,
The Journal of Mathematical Behavior, Volume 36, 2014, Pages 20-32,
https://doi.org/10.1016/j.jmathb.2014.08.003.

\end{thebibliography}

\clearpage

\begin{center}
\Huge Supplementary Material
\end{center}

\appendix
\setcounter{equation}{0}

\section{Enumeration example \label{enum_example}}

\renewcommand{\theequation}{\thesection.\arabic{equation}}

Detailed example of formula enumeration algorithm. Top-down base system (see main text) has been
used for simplicity. Ternary base digits were assigned to three RPN
calculator buttons as: 0 $\to$ E ($e$), 1 $\to$ LOG ($\log_x{y}$),
2 $\to$ POW ($x^y$). After code length 3 invalid codes were omitted to save space. 

\begin{tabular}{ccccc}
Enum & CODE & Syntax & RPN sequence & formula \\
0	&	0	    & VALID		& E 	& $e$\\
1	&	1	    & INVALID   &&\\ 
2	&	2	    & INVALID   &&\\ 
3	&	00	    & INVALID   &&\\ 
4	&	10	    & INVALID   &&\\ 
5	&	20	    & INVALID   &&\\ 
6	&	01	    & INVALID   &&\\ 
7	&	11	    & INVALID   &&\\ 
8	&	21	    & INVALID   &&\\ 
9	&	02	    & INVALID   &&\\ 
10	&	12	    & INVALID   &&\\ 
11	&	22	    & INVALID   &&\\ 
12	&	000	    & INVALID   &&\\ 
13	&	100	    & INVALID   &&\\ 
14	&	200	    & INVALID   &&\\ 
15	&	010	    & INVALID   &&\\ 
16	&	110	    & INVALID   &&\\ 
17	&	210	    & INVALID   &&\\ 
18	&	020	    & INVALID   &&\\ 
19	&	120	    & INVALID   &&\\ 
20	&	220	    & INVALID   &&\\ 
21	&	001	    & VALID	    &  E, E, LOG &	$\log_e{e}=1$\\
22	&	101	    & INVALID   &&\\ 
23	&	201	    & INVALID   &&\\ 
24	&	011	    & INVALID   &&\\ 
25	&	111	    & INVALID   &&\\ 
26	&	211	    & INVALID   &&\\ 
27	&	021	    & INVALID   &&\\ 
28	&	121	    & INVALID   &&\\ 
29	&	221	    & INVALID   &&\\ 
30	&	002	    & VALID		& E, E, POWER 	& $e^e$\\
31	&	102	    & INVALID   &&\\ 
32	&	202	    & INVALID   &&\\ 
33	&	012	    & INVALID   &&\\ 
34	&	112	    & INVALID   &&\\ 
35	&	212	    & INVALID   &&\\ 
36	&	022	    & INVALID   &&\\ 
37	&	122	    & INVALID   &&\\ 
38	&	222	    & INVALID   &&\\ 
\ldots &&&&\\
210	&	00101	& VALID		& E, E, LOG, E, LOG & $\log_e{1} = 0$ \\
219	&	00201	& VALID		& E, E, POWER, E, LOG & 	$\log_e{e^e} = e$\\
228	&	00011	& VALID		& E, E, E, LOG, LOG  &	$nan$\\
255	&	00021	& VALID		& E, E, E, POWER, LOG 	& $\log_{e^e}{e} = 1/e$\\
291	&	00102	& VALID		& E, E, LOG, E, POWER & $e^1 = e$\\
300	&	00202	& VALID		& E, E, POWER, E, POWER & $e^{(e^e)}$ 	\\
309	&	00012	& VALID		& E, E, E, LOG, POWER & $1^e = 1$\\
336	&	00022	& VALID		& E, E, E, POWER, POWER &  $(e^e)^e = e^{e^2}$\\
\ldots &&&&
\end{tabular}

\section{Enumeration of integer and rational numbers \label{rat}}

 If we restrict ourselves to simplest case of integer and rational numbers, enumeration procedure
 without repetition, i.e., one-to-one mapping of non-negative integers $i$ into integers $j$ is known:
 $$
 j = \frac{1}{4} (-1)^i \left( -2 i+(-1)^i-1 \right).
 $$
 
 Positive rationals $r$ can be enumerated by repeated composition of the function:
 $$
 \nxt{(r)} = \frac{1}{1 + 2 \lfloor r \rfloor  - r },
 $$
 starting with zero. 

Surprisingly, function $\nxt$ is composition of two \textit{self-inverse} functions:
$$
\inv(x) = \frac{1}{x}, \quad \ladder{(x)} = 1 + 2 \lfloor x \rfloor  - x. 
$$

\begin{figure}
\includegraphics[width=\textwidth]{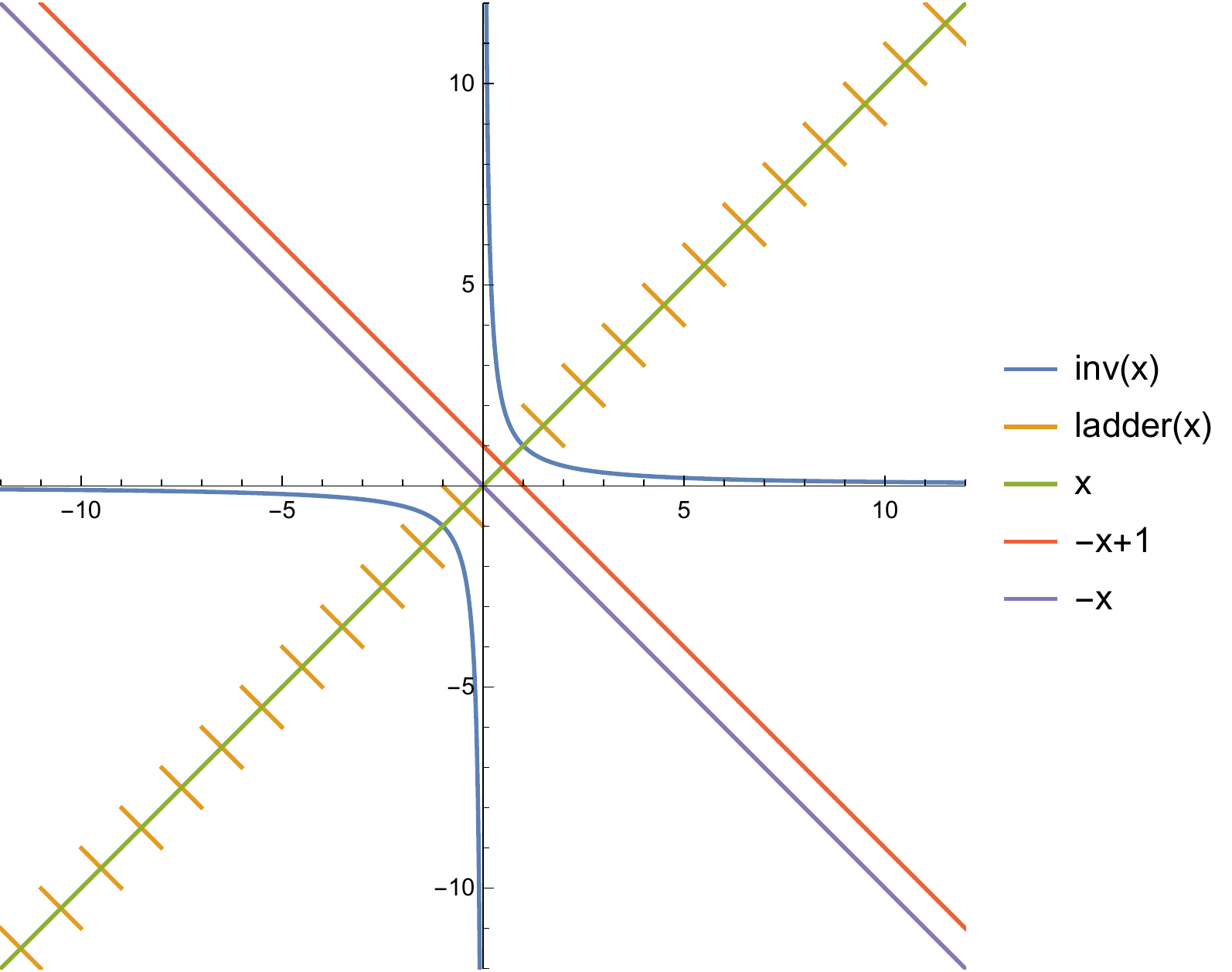}
\caption{\label{SelfInv} Self-inverse function, able to generate all integers and rationals without repetitions.
}
\end{figure}

Illustrative example is provided as follows. Let us define two additional self-inverse functions:
\begin{equation}
\minus(x)  = -x, 
\end{equation}
and
\begin{equation}
\pre(x)  = -x + 1.
\end{equation}

Any integer and rational (including negative) can be obtained be repeated composition of functions $\inv, \minus$ and $\pre$,
cf. Fig.~\ref{SelfInv}, starting with number (or a function) zero. Moreover, the appear in non-repetitive order:
$$
0,1, \infty, -1,2,\frac{1}{2},-2,3,-\frac{1}{2},\frac{3}{2},\frac{1}{3},
   -3,\frac{2}{3},4,-\frac{1}{3},-\frac{3}{2},\frac{4}{3},\frac{5}{2},\frac{1}{4},-\frac{2}{3},
   -4,\frac{3}{4},\frac{5}{3},5,\frac{2}{5},-\frac{1}{4},-\frac{4}{3},-\frac{5}{2}, \ldots
$$
 
Above properties are remarkable, and suggest possible way to non-repetitive generation of $\exp-\log$ numbers
by function composition, in  unique order. However, it is unclear what kind of function(s) is could be. For example, self-inverse function related to exponentiation is:
$$
e^{-1/\ln{x}},
$$
where -1 can be replaced by any other constant. In fact, it could be generalized to:
$$
p^{q/\log_p{x}},
$$
with arbitrary $p,q$. So far, our attempts to find $p,q$ failed. We only guess they are somehow related
to $e, \pi$ and possibly $i$.

\section{Proof of completeness of up-bottom base set \label{base-3-proof} }

Goal of his section is to show, that all explicit elementary complex numbers can be reduced to three
elements. 

We start with symbols:
$$
e, x^y, \log_x{y}.
$$

One can compute:
$$
\ln{x} = \log_e{x}, \quad 1=\ln{e}, \quad 0=\ln{1}, \quad 1/e = \log_{e^e}{e}, -1 = \ln{\log_{e^e}{e}}, 2 = \ln{\ln{(e^e)^e}}.
$$

Reversing role of logarithm base and argument, we also get:
$$
1/2 = \log_{e^{e^2}}{e}.
$$

This way one can compute all natural numbers and their reciprocals (egyptian fractions).

Multiplication can be computed by:
$$
x \cdot y = \log_{x}{ \left[ (x^y)^x  \right]}, 
$$
while division is:
$$
\frac{x}{y} = \log_{x^y}{x^x}.
$$

Doing addition is tricky, but possible:
$$
x+y = \log_x \; \log_{x^x} \left( \left(   (x^x)^{x^y} \right)^{x^x}  \right).
$$

Reciprocal is:
$$
1/x = \log_{x^x}{x},
$$
and sign change:
$$
-x = \log_x { \log_{x^{x^x}}{x} }.
$$

\begin{figure}
\includegraphics[width=0.5\textwidth]{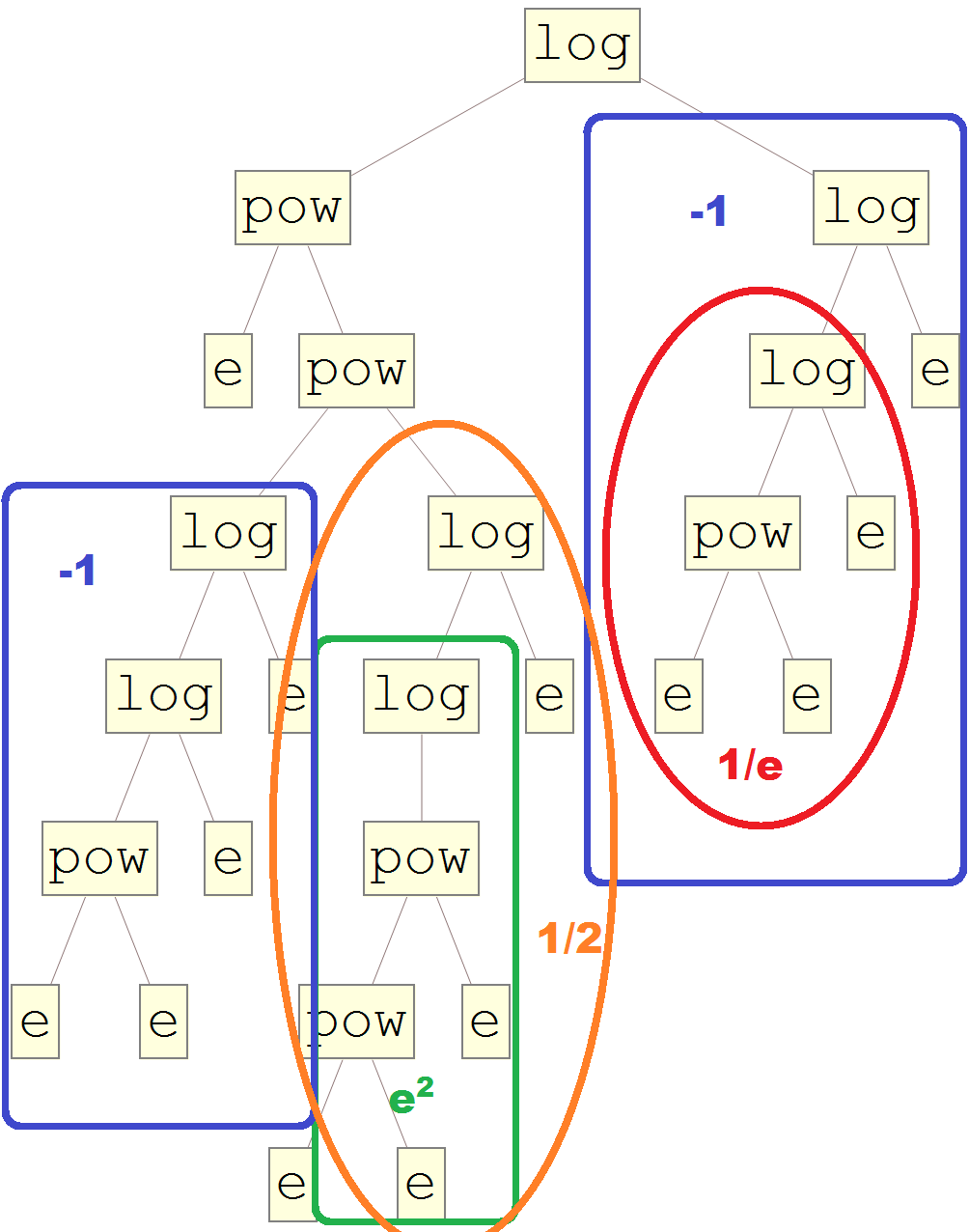}
\caption{\label{pi} Tree form of $\pi$ computed using primitive calculator CALC1.
}
\end{figure}

This complete basic 6 binary operations. We need only square root:
$$
\sqrt{x} = x^{\log _{\log _x\left(\left(x^x\right)^x\right)}(x)}
$$
of -1:
$$
i = \log_x\left(\log _{x^x}(x)\right)^{\log _{\log _x\left(\left(x^x\right)^x\right)}(x)},
$$
to compute remaining trigonometric functions. In particular (see Fig.~\ref{pi} for more readable form): 
$$
\pi = \log \left(e^{\log \left(\log \left(e^e,e\right),e\right)^{\log \left(\log
   \left(\left(e^e\right)^e\right),e\right)}},\log \left(\log
   \left(e^e,e\right),e\right)\right).
$$

This completes the proof.

\section{Proof of completeness of bottom-up base set \label{base-4-proof} }

Goal of his section is to show, that all explicit elementary complex numbers can be reduced to calulator with four
buttons.

We start with symbols:
$$
\exp, \ln, x, -.
$$

One can calculate:
$$
0 = x-x, \quad 1 = \exp{0} \equiv \exp{(x-x)}, \quad  e=\exp{(1)} \equiv \exp{(\exp{(x-x)})}.
$$

Now, we know how to compute:
$$
\exp, \ln, x, -, 0, 1, e.
$$

Addition is:
$$
x+y = x-(0-y) = x-((x-x)-y).
$$

One can change sign and compute reciprocal with:
$$
-x = 0-x = (x-x)-x, \quad 1/x = \exp{(-\ln{x})}.
$$
So far we have:
$$
\exp, \ln, x, -, 0, 1, e, +, 1/x,
$$
succesor is:
$$
x+1 = x-(0-1) = x - \left( (x-x) - \exp{(x-x)}  \right).
$$

Using succesor and reciprocal on can compute all integers and rationals. Multiplication
and division using logarithms are well-known:
$$
x \cdot y  = \exp{(\ln{x} + \ln{y})}, \quad  \frac{x}{y} = \exp{(\ln{x} - \ln{y})}.
$$

Binary exponentiation and logarithm are:
$$
x^y = \exp{(y \cdot \ln{x})}, \quad  \log_x{y} = \frac{\ln{y}}{\ln{x}}.
$$

Let's proceed to square root and $i$:
$$
\sqrt{x} = x^{1/2}, \quad i = \sqrt{-1}. 
$$

Number $\pi$ is:
$$
\pi = -i \ln{(-1)}.
$$

Now, calculating trigonometric functions is straightforward:
$$
\sin{x} = \frac{\exp{(i x)} - \exp{(-i x) }}{2 i}, \ldots
$$

Above shows, that all constants, functions and binary operations from Sect.~2 can be computed using CALC2.

\section{Distribution of EL numbers on complex plane \label{complex}}

\begin{figure}
\includegraphics[width=0.5\textwidth]{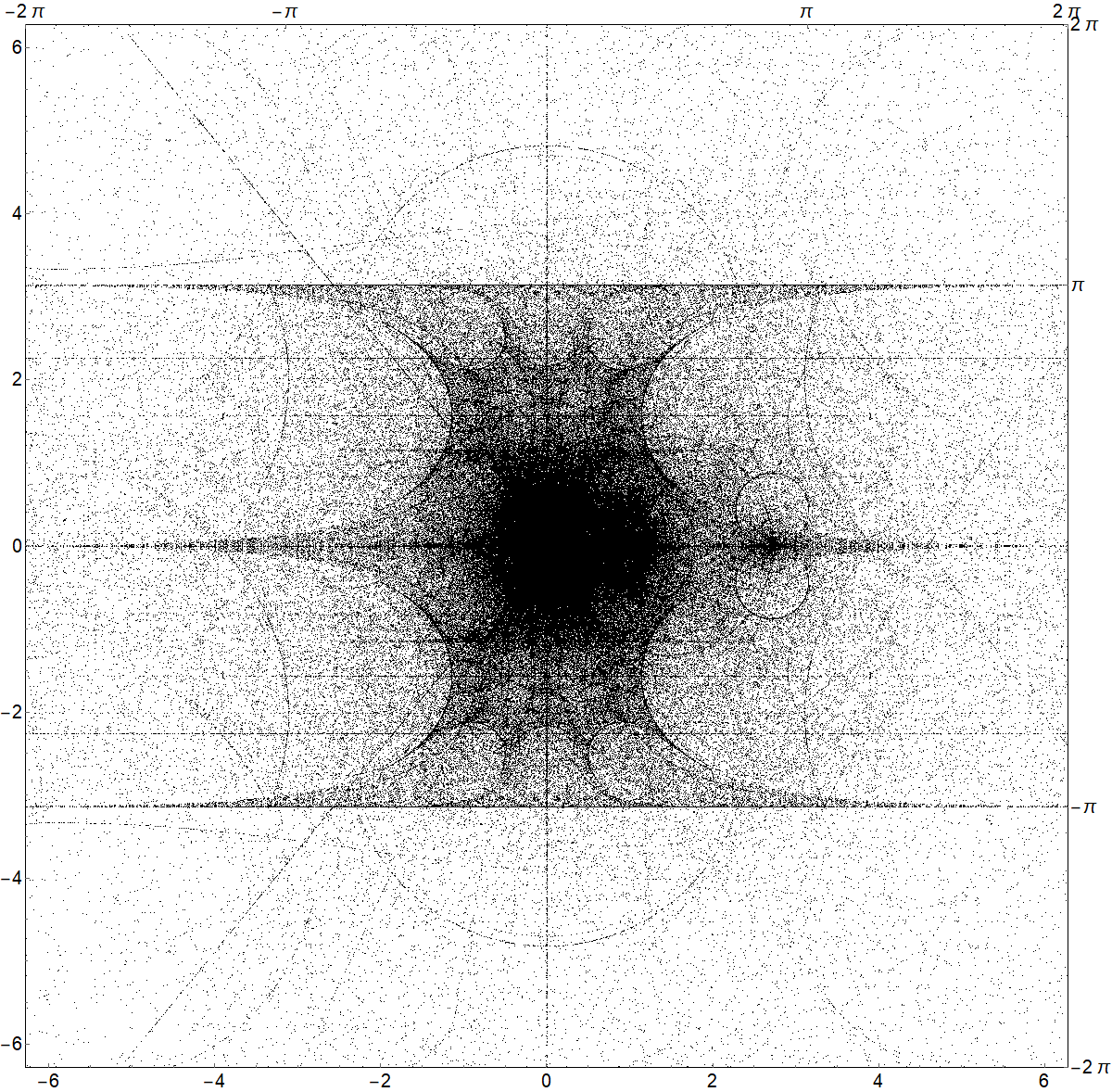}~\includegraphics[width=0.5\textwidth]{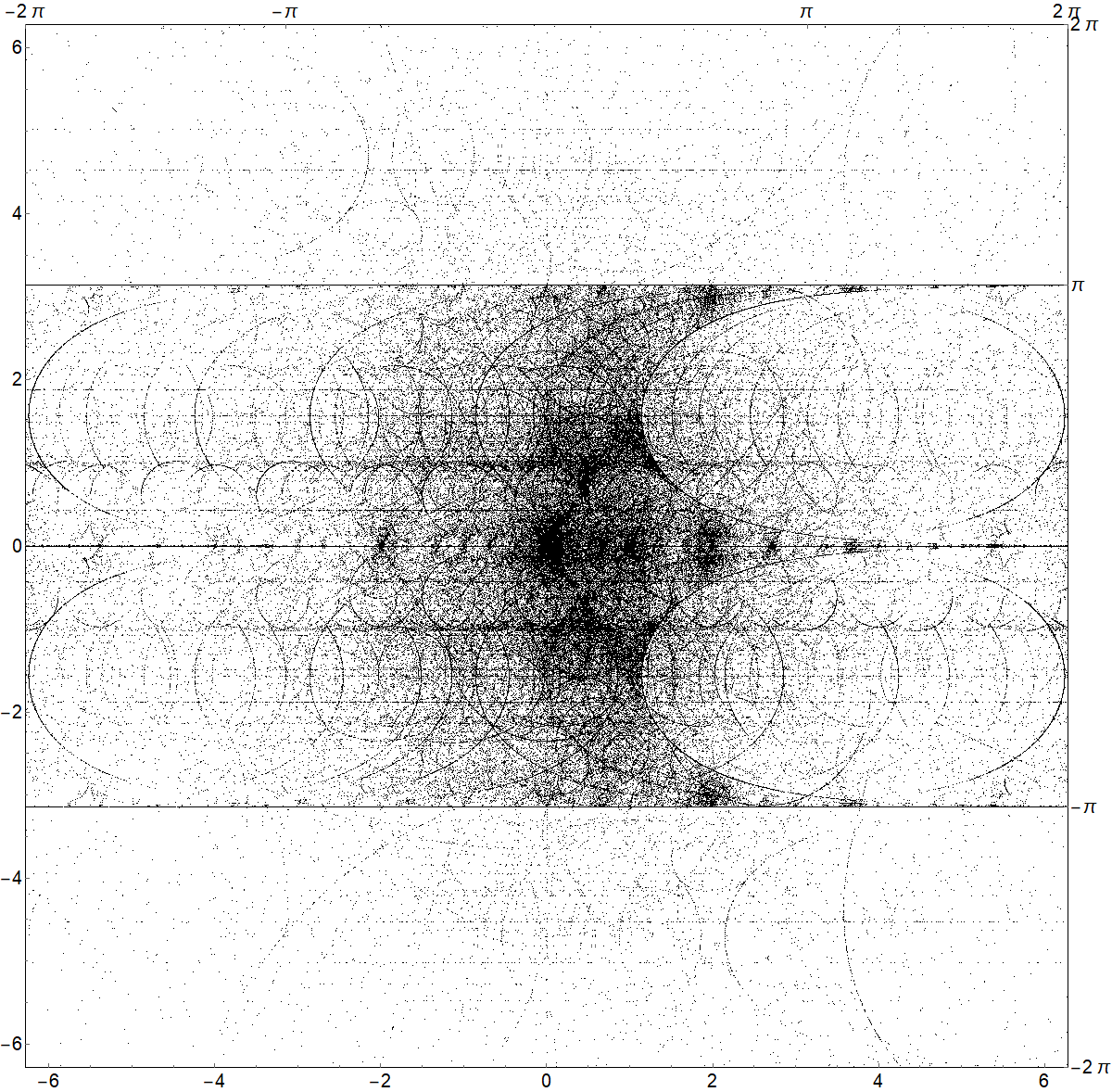}\\
\includegraphics[width=0.5\textwidth]{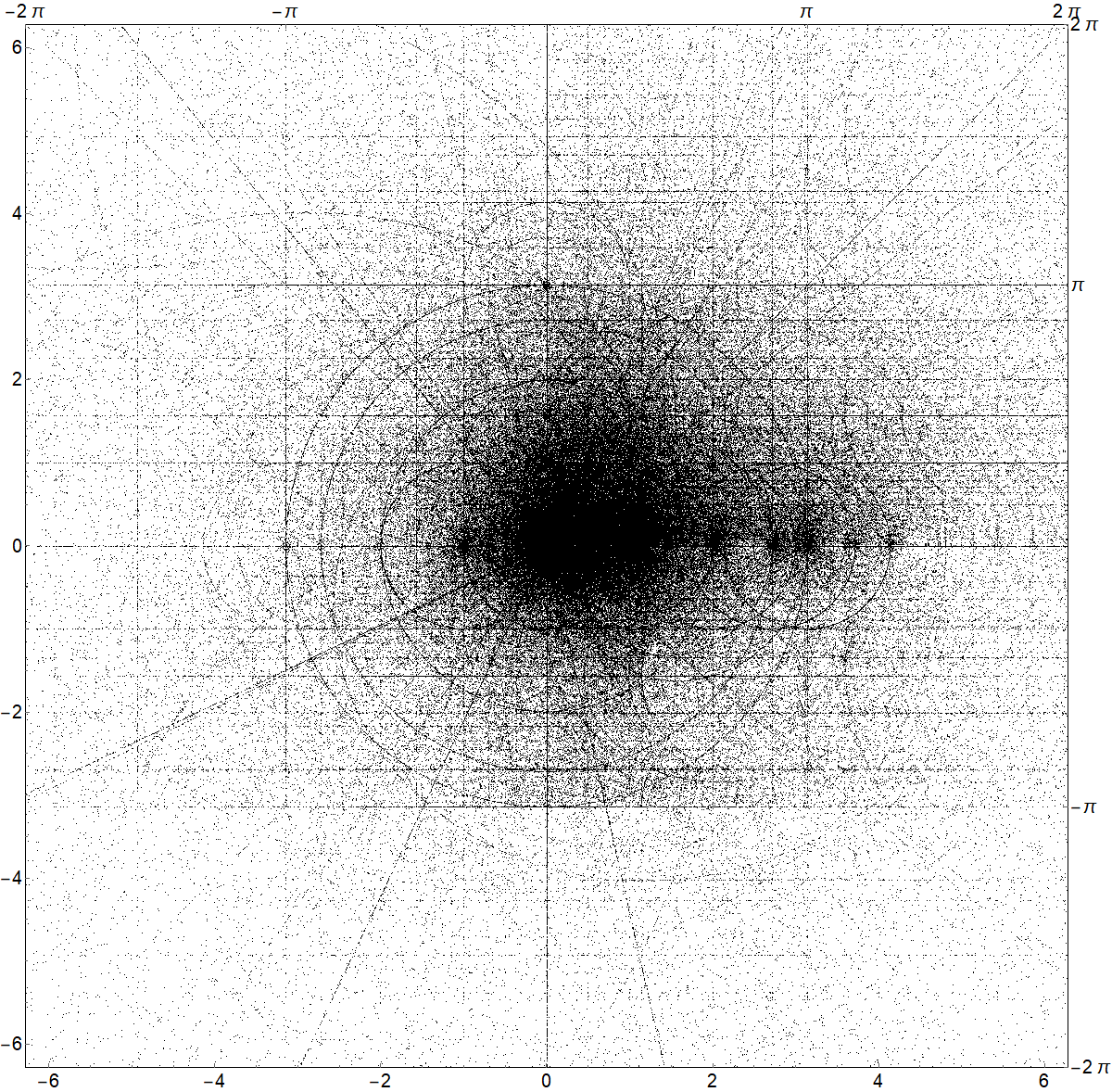}~\includegraphics[width=0.5\textwidth]{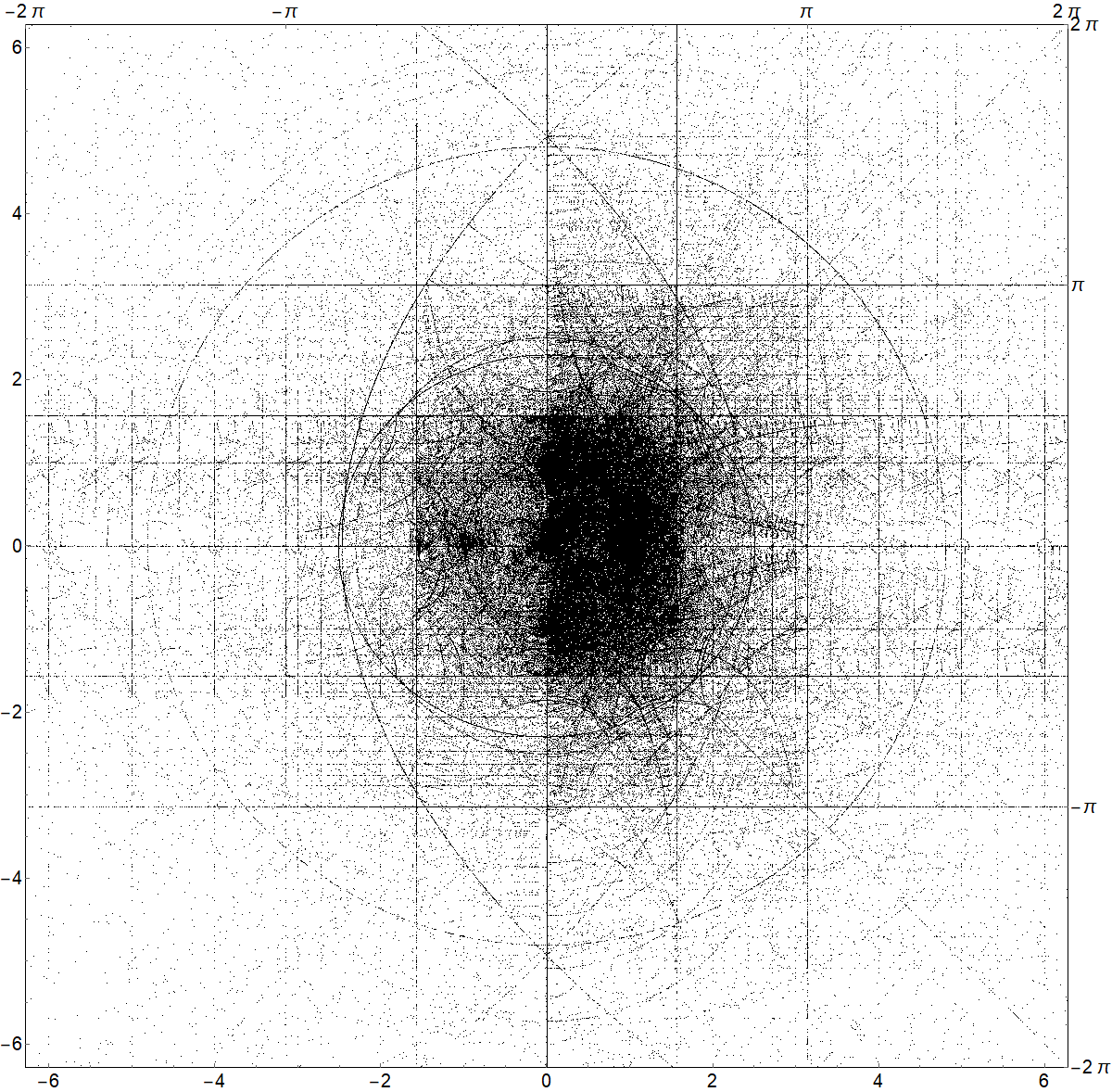}
\caption{\label{ComplexPlane} Distribution of the EL numbers on complex plane for calculators 1-4.
Calculator 1 (top-down) is shown in the upper-left panel. Only formulas with $\operatorname{Im}(z) \neq 0$ are shown,
up to Kolmogorov complexity $K \leq 23$, what gives 1500k unique complex numbers. Upper-right panel is for CALC2 (down-top),
up to $K \leq 16$; lower-left shows CALC3 (Mathematica) for $K \leq 9$; lower-right present full calculator CALC4 up to $K \leq 5$. 
}
\end{figure}

\begin{figure}
\includegraphics[width=0.5\textwidth]{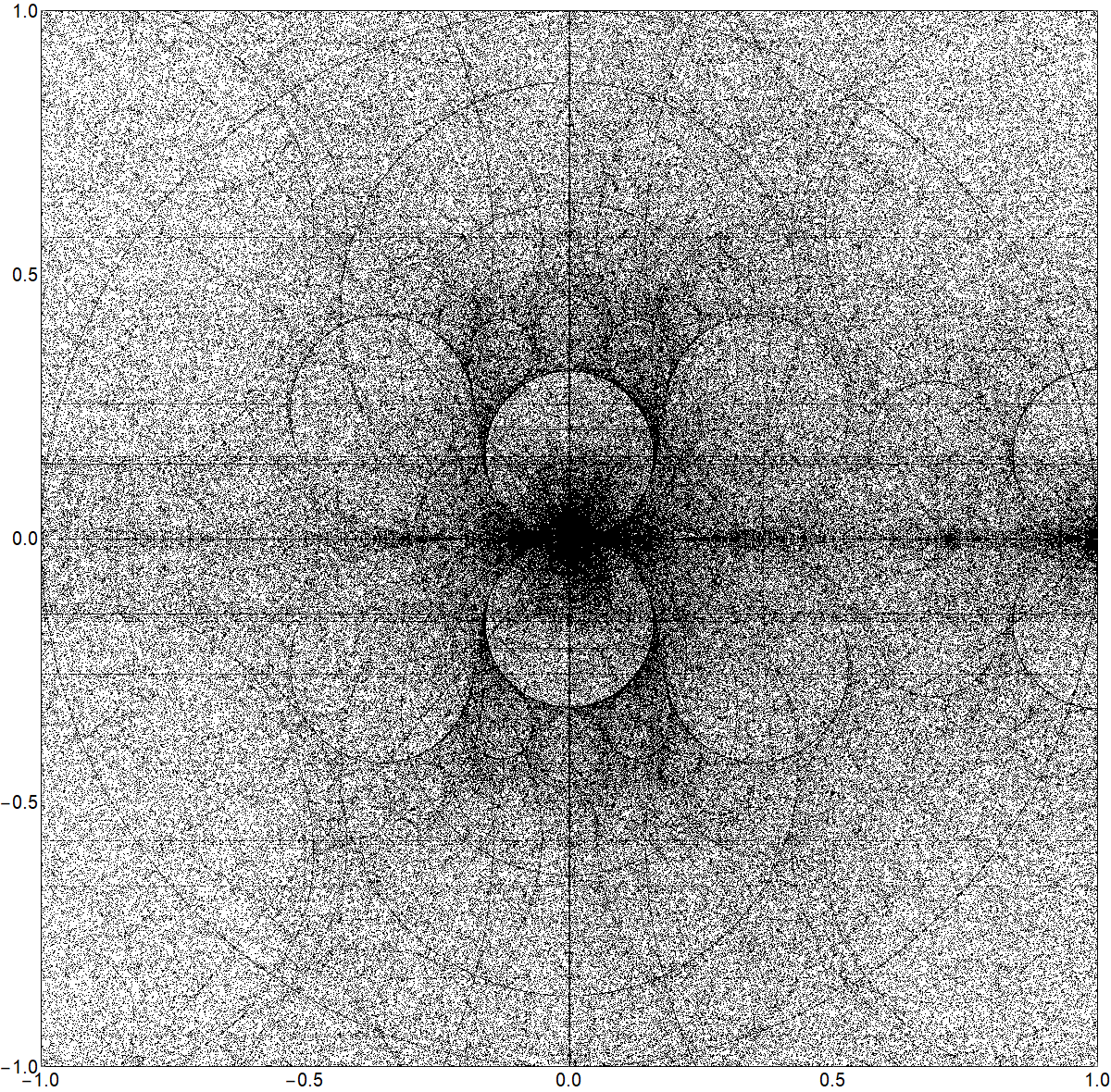}~\includegraphics[width=0.5\textwidth]{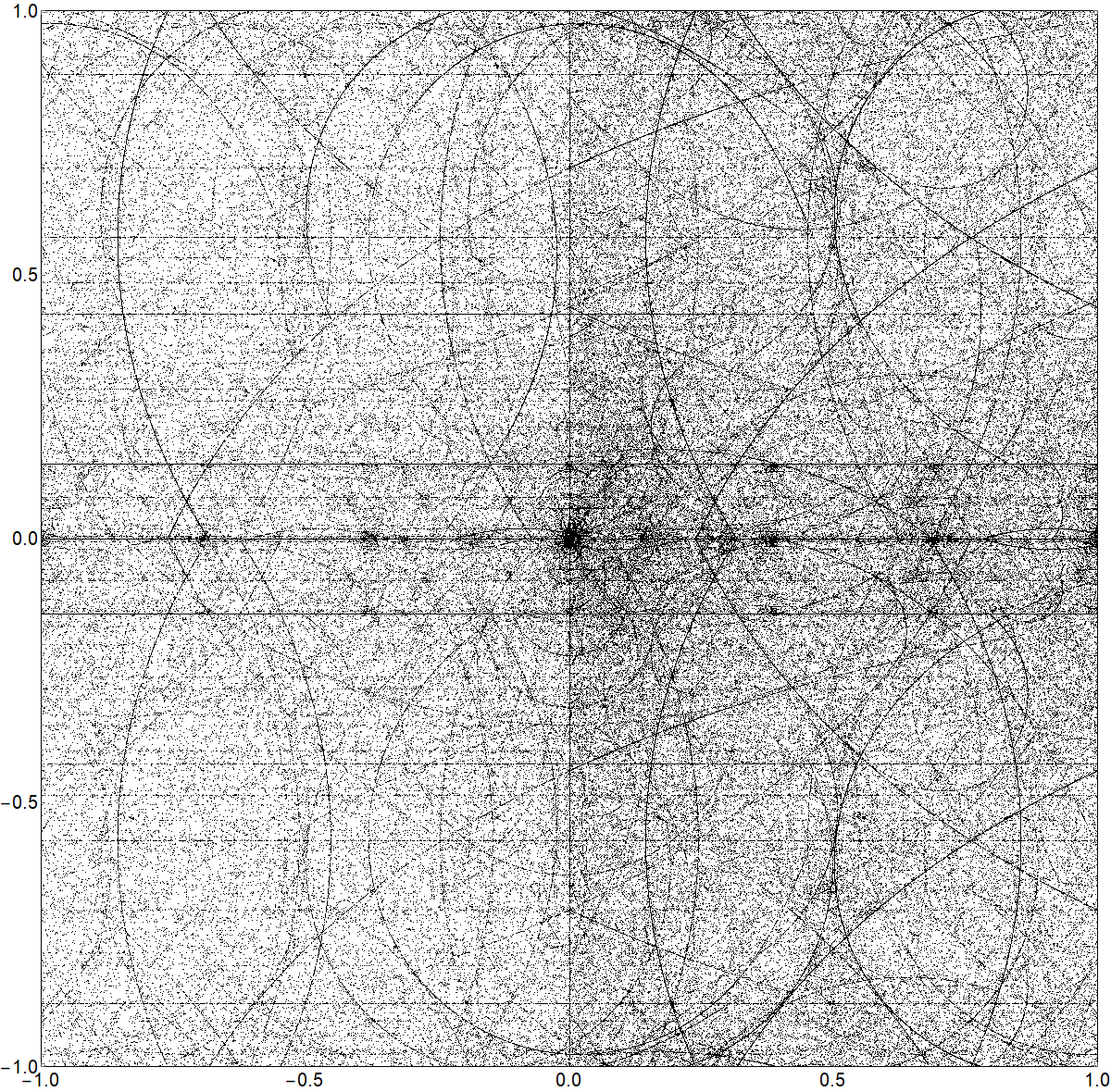}\\
\includegraphics[width=0.5\textwidth]{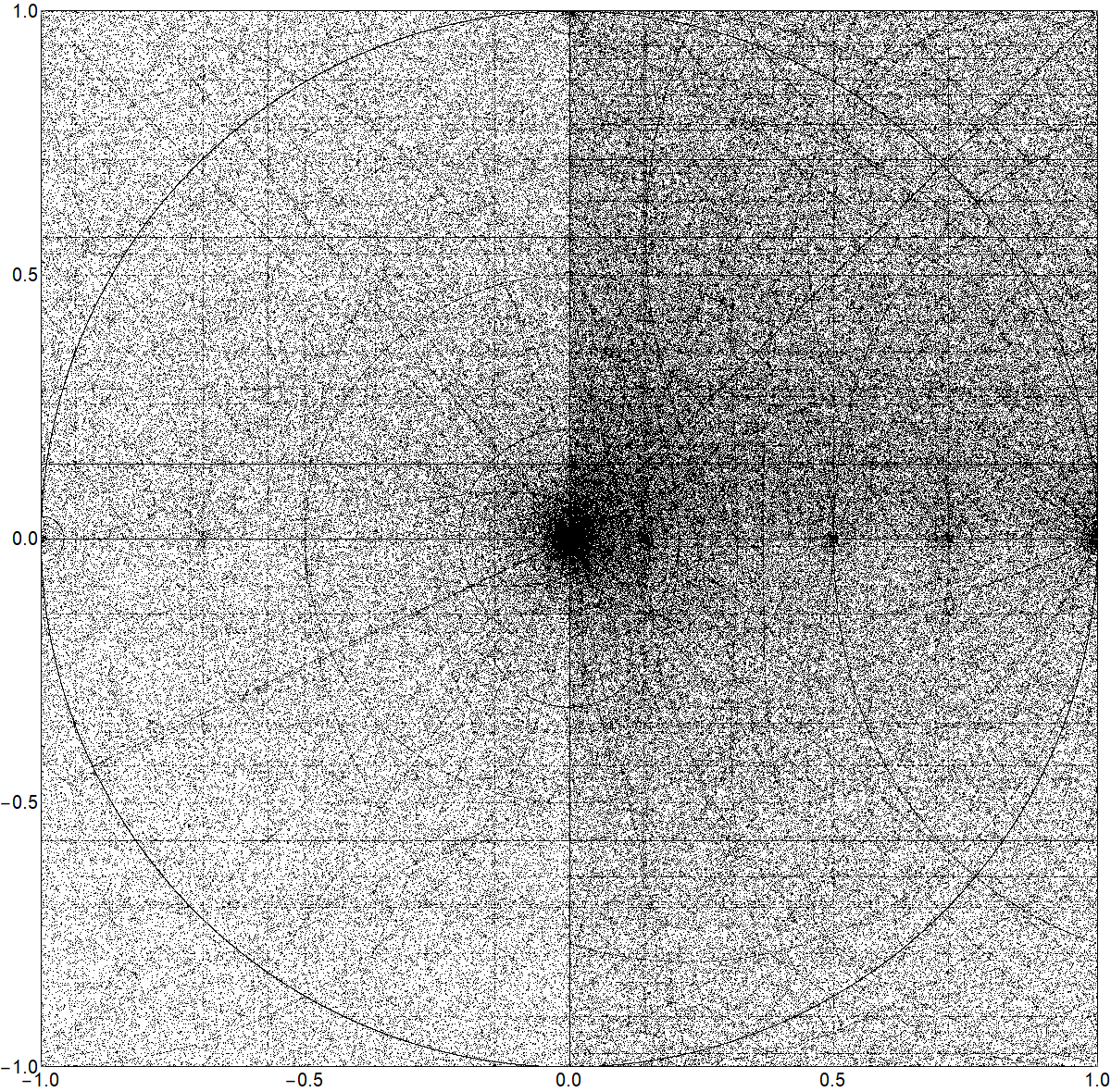}~\includegraphics[width=0.5\textwidth]{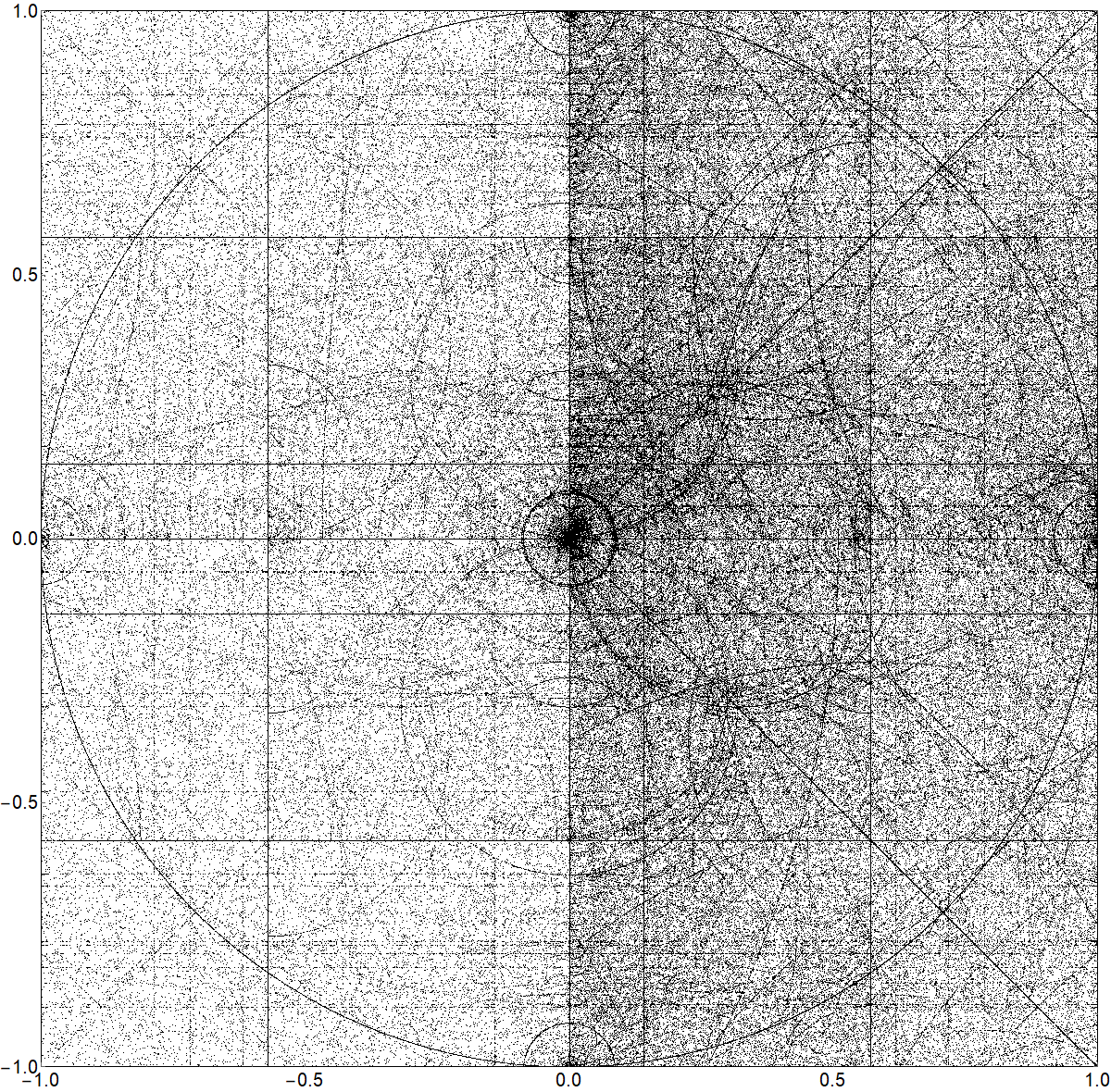}
\caption{\label{ComplexPlane_FractionalPart} Same as in Fig.~\ref{ComplexPlane}, but his time distribution of the fractional part of the EL numbers on complex plane is shown.
}
\end{figure}

Distrubution of the EL numbers generated by sequence on complex plane is presented in Figs.~\ref{ComplexPlane}
and \ref{ComplexPlane_FractionalPart}. Visually, it is far from random. However, it is not fractal, a because
EL numbers include rationals, which are everywhere dense.

\end{document}